# Galactic Chemical Evolution and the Oxygen Isotopic Composition of the Solar System

Larry R. Nittler[a] and Eric Gaidos[b]

[a]Department of Terrestrial Magnetism, Carnegie Institution of Washington, Washington, DC 20015, USA lnittler@ciw.edu

[b]Department of Geology & Geophysics, University of Hawaii at Manoa, Honolulu, HI 96822, USA, gaidos@hawaii.edu

# Abstract

We review current observational and theoretical constraints on the Galactic chemical evolution (GCE) of oxygen isotopes in order to explore whether GCE plays a role in explaining the lower $^{17}O/^{18}O$ ratio of the Sun, relative to the present-day interstellar medium, or the existence of distinct $^{16}O$-rich and $^{16}O$-poor reservoirs in the Solar System. Although the production of both $^{17}O$ and $^{18}O$ are related to the metallicity of progenitor stars, $^{17}O$ is most likely produced in stars that evolve on longer timescales than those that produce $^{18}O$. Therefore the $^{17}O/^{18}O$ ratio need not have remained constant over time, contrary to preconceptions and the simplest models of GCE. An apparent linear, slope-one correlation between $\delta^{17}O$ and $\delta^{18}O$ in the ISM need not necessarily reflect an O isotopic gradient, and any slope-one galactocentric gradient need not correspond to *evolution in time*. Instead, increasing $^{17}O/^{18}O$ is consistent both with observational data from molecular clouds and with modeling of the compositions of presolar grains. Models in which the rate of star formation has decelerated over the past few Gyr or in which an enhanced period of star formation occurred shortly before solar birth ("starburst") can explain the solar-ISM O-isotopic difference without requiring a local input of supernova ejecta into the protosolar cloud. "Cosmic chemical memory" models in which interstellar dust is on average older than interstellar gas predict that primordial Solar System solids should be $^{16}O$-rich, relative to the Sun, in conflict with observations. However, scenarios in which the $^{16}O$-rich contribution of very massive stars could lead to $^{16}O$-poor solids and a $^{16}O$-rich bulk Sun, if the Solar System formed shortly after a starburst, independent of the popular scenario of photochemical self-shielding of CO.

# Introduction

The Sun formed some 9 Gyr after the Big Bang and its elemental and isotopic composition is a product of billions of years of intervening Galactic evolution. Models of Galactic Chemical Evolution (GCE) attempt to explain variations in the elemental and isotopic composition with position and time due to succeeding generations of stars ejecting newly synthesized elements into the interstellar medium (ISM). Oxygen is the third most abundant element in the Universe after H and He and is present in both the gas and solid phases of the ISM and solar nebula; its elemental abundance can be readily determined in solar-type stars and its isotopes ($^{16}$O, $^{17}$O, $^{18}$O) can be measured (as isotopomers of CO, OH, and $H_2CO$) in the ISM and cores of molecular clouds (MCs). The Sun is the best characterized of all stars and is commonly used as a benchmark for GCE models, based on the assumption that it is typical for its age and location in the Galaxy. In addition to chondritic meteorites, from which primordial solar system abundances can be inferred, there are now *Genesis* measurements of the solar wind (McKeegan et al., 2011).

GCE models can reproduce the observed age-oxygen abundance relation of solar-mass stars and the approximate proportionality between the secondary isotopes $^{17}$O and $^{18}$O, both of which should by synthesized in amounts proportional to the primary isotope $^{16}$O. However, there are two outstanding enigmas involving the O isotopic composition of our Solar System. First, the present-day ISM has a significantly lower $^{18}$O/$^{17}$O ratio (~4) than that of the Solar System (5.2). This has been interpreted as evidence for an enrichment of the protosolar molecular cloud by $^{18}$O-rich supernova ejecta before or during the formation of the Solar System (Young et al., 2011). This offset may test models of the setting and mechanism of the formation of the Sun from its parent molecular cloud.

Second, oxygen isotopes are heterogeneously distributed in the Solar System, with Earth, Moon, Mars, asteroids, and cometary dust systematically depleted in $^{16}$O by several percent relative to the Sun. The most popular explanation for this difference is that it was caused by the UV dissociation of CO (and self shielding of abundant $C^{16}O$) either in the Sun's parental molecular cloud (Yurimoto and Kuramoto, 2004) or in the solar nebula (Lyons and Young, 2005), chemical separation into two isotopically fractionated reservoirs, and later inhomogenous mixing. In this interpretation, the isotopic difference is informing us of the UV radiation environment and photochemistry of the protosolar cloud or disk. Alternatively, the isotopically distinct reservoirs may have been primordial (i.e. pre-existing in the parent cloud), and perhaps the outcome of GCE for solid and gas phases of the ISM (Clayton, 1988;

Krot et al., 2010). Note that because this solar-planetary variability is primarily in the abundance of $^{16}$O, the $^{18}$O/$^{17}$O ratio of the Solar System remains well defined ($^{18}$O/$^{17}$O = 5.2 ± 0.2, Young et al., 2011).

Clearly, understanding O-isotopic ratios in the context of GCE is critical to understanding where and from what material the Solar System formed. In this paper, we discuss the current state of knowledge of the Galactic evolution of oxygen isotopic ratios from both an observational and theoretical perspective, and address questions of how the Solar System obtained its O-isotope composition and the role played by global GCE processes in the galactic disk versus 'local' processes, e.g., SN enrichment of the protosolar cloud during sequential star formation. Specifically, we seek answers to the questions: do current observations support deviations from a constant $^{18}$O/$^{17}$O ratio during GCE? Can the distinct $^{18}$O/$^{17}$O ratio of the Sun compared to the present-day ISM be explained by GCE or does it require SN enrichment of the protosolar cloud? Is there a 'Galactic' explanation for the different O-isotopic compositions of the Sun and planets?

Realistic models of GCE include infall of primordial gas, appropriate treatment of star formation, and adequate resolution of the stellar mass function. The approach we adopt here is to use the simplest non-trivial models to explore which effects are most important in the GCE of oxygen. We caution that all GCE models have substantial uncertainties and a model is often considered successful even it only approximately reproduces observations , i.e. the metallicity distribution of low-mass stars and the solar composition at the epoch of its formation, to within a factor of two. Predictions of relative isotopic and elemental abundances are susceptible to (sometimes enormous) uncertainties in nucleosynthesis yields (Romano et al., 2010). For example, the production of $^{17}$O from Type II supernovae predicted by the same group decreased by more than a factor of 10 as a result of new laboratory measurements of important reaction cross sections (Woosley and Heger, 2007; Woosley and Weaver, 1995, hereafter WW95). Therefore, one should treat specific quantitative predictions of models with great caution when confronting the high-precision O-isotopic measurements of materials in the Solar System.

# GCE of oxygen isotopes

**Basic Considerations**

Oxygen has three stable isotopes, $^{16}$O, $^{17}$O, and $^{18}$O, which are present on the Earth in fractional abundances of ~99.76%, 0.038% and 0.2%, respectively. In geochemistry and cosmochemistry, O-isotope ratios are commonly expressed as δ-values, fractional deviations from standard (normalizing) ratios, and data are often presented in a "three-isotope" plot, *e.g.,* with $δ^{18}$O on the abscissa and $δ^{17}$O on the ordinate. Delta-values can be defined either as linear (e.g., $δ^{18}O = 10^3 \times [(^{18}O/^{16}O)/(^{18}O/^{16}O)_0 – 1]$, where "0" denotes a normalization ratio) or logarithmic (e.g., $δ^{18}O' = 10^3 \times \ln[(^{18}O/^{16}O)/(^{18}O/^{16}O)_0])$ functions of the normalized isotope ratio. Logarithmic delta-values have the advantage that data with the same $^{18}$O/$^{17}$O ratios lie along a line of slope one in a three-isotope plot and were used by Young et al. (2011) in their discussion of O-isotope GCE and the solar composition. However, they are not commonly used in the astronomical literature, and for large deviations from zero their quantitative meaning is not always intuitively obvious. For example, a linear $δ^{18}$O value of –1000 corresponds to $^{18}$O/$^{16}$O=0, but a logarithmic $δ^{18}$O' value of –1000 means $^{18}$O/$^{16}$O=0.37 times the normalizing ratio. Thus, for clarity in this paper, where we present data as logarithmic delta values, we also show the normalized isotopic ratios without the natural logarithm applied.

Models of oxygen isotopic abundances in our galaxy have roots in nearly a half-century of theoretical studies of GCE (Audouze and Tinsley, 1976; Prantzos et al., 1996); one recent review is by Meyer et al. (2008). As stars evolve and contribute newly-synthesized nuclei to the ISM, the galactic inventories of heavy elements (metallicity) increase. However, because different isotopes are made by different nuclear processes in different types of stars, which may evolve on very different timescales, isotopic and elemental ratios also vary with metallicity. GCE theory distinguishes between *primary* $^{16}$O, whose synthesis in stars is independent of metallicity, and *secondary* $^{17}$O and $^{18}$O, which require preexisting C, N and/or O. Simple analytical GCE models (Clayton, 1988) predict that the ratio of a secondary to a primary isotope, e.g.,$^{17}$O/$^{16}$O, should increase linearly with metallicity, and the ratio of two secondary isotopes, e.g., $^{18}$O/$^{17}$O, should remain constant. In a three-isotope plot, the O-isotopic composition of the Galaxy should evolve along a slope-one line towards the upper right. This simplest analysis is based on the assumption that all three O isotopes are produced in short-lived massive stars. If this accurately describes GCE, then the average $^{18}O/^{17}O$ ratio of the ISM has not changed since 4.6 Gyr ago, and the unusual $^{18}O/^{17}O$ ratio of the Sun compared to the present-day ISM (Fig. 1), requires another explanation, e.g., local pollution of the Sun's parental cloud by the ejecta from one or more supernovae (Prantzos et al., 1996; Young et al., 2011).

**Astronomical Constraints**

Radio observations of molecular clouds in the Galaxy over the last few decades have supported the view outlined above of O-isotopic GCE (Penzias, 1981; Wannier, 1980; Wilson and Rood, 1994). Namely, there is evidence for gradients of increasing $^{17,18}O/^{16}O$ ratios with decreasing distance from the galactic center, but relatively constant $^{18}O/^{17}O$ ratios throughout the Galaxy. Since the overall metallicity of the Galaxy is observed to increase towards the galactic center, this supports the idea that $^{16}O$ is primary and $^{17}O$ and $^{18}O$ are secondary. However, the radio measurements entail relatively large uncertainties, especially for most measurements of $^{17}O/^{16}O$ and $^{18}O/^{16}O$, which require measuring double isotopic ratios (e.g., $^{12}C^{18}O/^{13}C^{16}O$) and making assumptions regarding $^{12}C/^{13}C$ ratios.

Figure 1 shows the $^{18}O/^{17}O$ ratios of molecular clouds throughout the Galaxy from the pioneering study of Penzias (1981) and the much more recent work of Wouterloot et al. (2008). Both studies are consistent with this ratio being basically constant across much of the Galaxy and lower on average than the well-defined solar ratio. Moreover, independent infrared measurements (Smith et al., 2009) of O-isotopes in two young stellar objects (also shown on Fig. 1) show that they have similar $^{18}O/^{17}O$ ratios to those derived from the radio data, demonstrating that the offset between the solar ratio and typical ISM compositions is not due to some systematic uncertainty in the reduction of the radio data. There is a systematic offset towards higher $^{18}O/^{17}O$ ratios in the more recent radio work, most likely reflecting refinement of observational and analysis techniques over several decades. The Wouterloot et al. (2008) data show that at least one MC at the same galactocentric distance as the Sun has an $^{18}O/^{17}O$ ratio close to solar, hinting that cloud-to-cloud scatter may be as important as overall average trends. More importantly, the Wouterloot et al. (2008) study finds significantly lower $^{18}O/^{17}O$ ratios than earlier work in the Sagittarius-B MC close to the Galactic center and higher ratios in the outer Galaxy. While clearly needing additional investigation, this suggestion of a radial gradient is qualitatively consistent with the evolution of the Galaxy towards more $^{17}$O-rich compositions with time.

To derive a galactic gradient in $^{18}$O/$^{16}$O ratio, Young et al. (2011) combined the $^{13}$C$^{16}$O/$^{12}$C$^{18}$O line ratios reported by Wouterloot et al. (2008) with an independent estimate of the galactic gradient in $^{12}$C/$^{13}$C ratios from Milam et al. (2005). Caution is of course necessary when combining disparate datasets, each with its own uncertainties and potential biases, in this way, as it can be very difficult to realistically estimate the uncertainty in the result. For example, Wouterloot et al. (2008) report $^{13}$C$^{16}$O/$^{12}$C$^{18}$O line ratios for two different CO rotational transitions (J=1-0 and J=2-1), and these datasets indicate significantly different galactic gradients in $^{18}$O/$^{16}$O even if combined with the same galactic $^{12}$C/$^{13}$C gradient (Fig. 2). In fact these authors state that chemical fractionation, not a trend in bulk galactic isotope composition, is a likely explanation for the observed $^{13}$C$^{16}$O/$^{12}$C$^{18}$O gradients.

In Fig. 2, we compare the galactic gradients in $^{16}$O/$^{18}$O derived as described above from the studies of Wouterloot et al. (2008) and Milam et al. (2005) with those compiled by Wilson and Rood (1994), from double-isotope measurements of formaldehyde, and measured by Polehampton et al. (2005), from OH measurements. The latter study is important as it is a direct measurement, not requiring a correction for any other isotopic ratios (i.e., $^{12}$C/$^{13}$C for the CO and H$_2$CO studies). The gradient derived from the CO J=1-0 data is clearly steepest, whereas the OH data show little evidence for any gradient at all. A striking feature of Fig. 2 is the large scatter in the data, both between different studies, and for different clouds within a single study. For example, the CO J=1-0 data show a factor of ~3 range in $^{16}$O/$^{18}$O ratio for clouds close to the solar galactocentric radius. Such scatter may be real; there is mounting evidence for large scatter in metallicity among stars of the same age in the solar neighborhood (Edvardsson et al., 1993; Nordström et al., 2004) and this could translate directly into O-isotopic variations assuming that the O-isotopic ratios scale with metallicity via GCE processes. We emphasize that a spatial gradient (with galactocentric radius) need not imply the same evolution in time: The ISM that we observe is everywhere comparatively "young" compared to stars and the Galaxy itself.

Moreover, the evident scatter in O-isotopic ratios for relatively nearby MCs has other important implications for our discussion. First, if it is real, it indicates that it is too simplistic to consider that a single O-isotopic composition pertains to even relatively nearby and contemporaneous regions of the Galaxy. Second, the data clearly show that the local ISM $^{16}O/^{17}O$ and $^{16}O/^{18}O$ ratios are highly uncertain by at least a factor of two, far greater than the ±20% assumed in the SN enrichment model of Young et al. (2011).

**Non-canonical models of GCE: a two phase ISM**

Nearly all models of O-isotopic GCE adopt the Instantaneous Mixing Approximation (IMA), which assumes that the ISM consists of a single homogeneous reservoir. The IMA cannot be absolutely correct; winds from AGB stars and SN ejecta are hot and tenuous compared to molecular clouds and will not mix efficiently; and SNe drive bubbles and "galactic fountains" that eject hot gas into the halo. Gilmore (1989) suggested that molecular clouds may be self-enriched by the very massive stars they spawn. Malinie et al. (1993) proposed that star formation was both heterogeneous and bursting (episodic), and different parts of the galactic disk mix only after 0.1-1 Gyr had elapsed. Thomas et al. (1998) relaxed the IMA, dividing the ISM into separate "active" (cool, dense, and star-forming) and "inactive" (hot, less dense) phases. Stellar ejecta first enter the inactive phase and only enter the "active" phase after a characteristic time $\tau_I$. They found that $\tau_I = 0.1$ Gyr gave only a slight improved description of the metallicity evolution of the galactic disk compared to an IMA, and only for the first few Gyr. A value of $\tau_I = 1$ Gyr was ruled out by the existence of very old metal-rich stars. Spitoni et al. (2009) also examined time delays of up to 1 Gyr due to the ejection of gas in galactic "fountain" and the time for it to return by cooling. They found that such delays have only a small effect on elemental chemical evolution and radial abundance gradients. They also found that the use of highly metallicity-dependent yields of WW95 does induce an effect, but this choice also predicts elemental ratios (i.e. [O/Fe]) that are contradicted by observations. Note that the IMA is independent of assumptions concerning radial transport and mixing in the galactic disk (Schönrich and Binney, 2009).

More subtle influences of heterogeneity and non-instantaneous mixing might be detected in O-isotopic compositions. Gaidos et al. (2009) used a two-phase model analogous to Thomas et al. (1998) and consisting of a diffuse ISM in which no star formation occurs, plus star-forming molecular clouds. Their model produces two effects: an O-isotopic offset between the two reservoirs, and trajectories in a three-isotope plot that deviated from the canonical slope-one line, with a general trend to lower $^{18}O/^{17}O$ ratios with time. The latter is explained both as a result of depletion of the ISM of $^{18}O$ as a result of its preferential incorporation into low-mass stars born in self-enriched star-forming regions, and preferential injection of $^{17}O$-rich AGB ejecta into the diffuse ISM. To more fully illustrate and explore the outcomes of these effects on solar and ISM $^{18}O/^{17}O$ ratios, we consider here some simple two-phase ISM models of GCE.

We model the ISM as consisting of two phases: a diffuse, inactive ISM (phase "I") and denser, star-forming regions (SFRs, or phase "S"). No mass is lost, except through the formation of long-lived stars and stellar remnants. In steady-state this loss is balanced by the infall of primordial, metal-free gas into the diffuse ISM. The governing equations for nucleosynthetic isotopes in this two-phase model are

$$\Sigma_I \frac{dx_i}{dt} = p_i - Fx_i + Dy_i, \qquad (1)$$

and

$$\Sigma_S \frac{dy_i}{dt} = q_i + Fx_i - (D+R)y_i \qquad (2)$$

where $\Sigma$ is a mass surface density (constant in steady-state), $x_i$ and $y_i$ are the mass fractions of the $i^{th}$ isotope in the diffuse ISM and SFRs, respectively, $p_i$ and $q_i$ are the respective rates of injection of the isotope by both SN and AGB stars, $F$ and $D$ are the rates of formation and dissipation of SFRs, and $R$ is the total rate of star formation. This model differs from that of Thomas et al. (1998) in that SN ejecta is assumed to be first added to the residual gas of SFRs before it is dispersed back into the diffuse ISM.

In steady-state, mass balance in the SFR population requires $F = D + R$. If we assign a characteristic lifetime $\tau_S$ to SFRs before they are recycled back into the diffuse ISM, then $D = \Sigma_S/\tau_S$ and $F = \Sigma_S(1+s)/\tau_S$ where $s = R\tau_S/\Sigma_S$ is the star formation efficiency. Then,

$$(1+s)x_i - y_i = p_i \tau_S/\Sigma_S, \qquad (3)$$

$$-(1+s)x_i - (1+s)y_i = q_i \tau_S/\Sigma_S, \qquad (4)$$

to which the solutions are;

$$x_i = \frac{p_i + q_i/(1+s)}{R},\qquad(5)$$

and

$$y_i = \frac{p_i + q_i}{R}.\qquad(6)$$

Thus, in steady state the isotopic composition of SFRs is displaced by an amount $q_i s/[R(1+s)]$ with respect to the diffuse ISM. This offset represents contamination by stars that inject isotopes into SFRs after a lifetime shorter than $\tau_S$. If $\tau_S = 10$ Myr (Williams et al., 2000), this includes O stars more massive than 15 $M_\odot$. Such massive stars produce comparatively little $^{17}$O and $^{18}$O and thus the displacement of the SFR trajectory relative to the ISM trajectory in a three-isotope plot is primarily along the slope-one line. The low efficiency of star formation (s ~ 0.1) (Williams et al., 2000) also means that the amplitude of the displacement is small. This solution essentially assumes that the *typical* SFR undergoes the local SN enrichment invoked by Young et al. (2011) to explain the anomalous O-isotopic composition of the Sun, but as we shall see, the average effect on $^{18}$O/$^{17}$O is smaller than that predicted by those authors. Is this solution stable? The eigenvalues of the system of differential equations 1 and 2 are

$$\lambda = -\frac{(1+s)(1+m)}{2\tau_S}\left[1\pm\sqrt{1-\frac{4mf}{(1+s)(1+m)^2}}\right]\qquad(7)$$

where $m = \Sigma_S/\Sigma_I = 1$. They are real and negative for all values of $m$ and $s$, thus the steady-state solution is stable; transients decay on short ($\tau_S$) and long ($\tau_I \approx \tau_S/(mf)$) timescales.

Figure 3 plots solutions using IMF-integrated yields of oxygen isotopes for four different values of metallicity. The assumption of quasi-steady-state holds if the changes in metallicity or star formation rate are small compared to the lifetime of the longest-lived significant contributors to O inventories (perhaps ~2 Gyr) and must break down at early epochs and very low metallicities (see below). For the adopted set of yields (listed in the caption to Fig. 3), the predicted $^{18}$O/$^{16}$O ratio at solar metallicity is within 7% of the solar value; however, $^{17}$O is over-predicted by a factor of two. (Kobayashi et al. 2011, hereafter K11, arrived at a like result with a similar set of yields). A discrepancy of this magnitude is often assigned to uncertainties in the yields of $^{17}$O. The predictions are plotted as logarithmic δ-values normalized to the calculated composition of solar-metallicity SFRs. This procedure essentially normalizes out the uncertainty in the absolute predictions and allows relative trends to be examined (Timmes and Clayton, 1996). The main effect is that SFRs are $^{16}$O-enriched with respect to the diffuse ISM (inset of Fig. 3), and the diffuse ISM is only slightly (~1.5 %) $^{17}$O-rich compared to SFRs. The latter difference is far smaller than the observed difference in $^{18}$O/$^{17}$O between the Solar System (5.2) and ISM (~4) and stands in contrast with explanations based on self-contamination in OB associations (e.g., Prantzos et al., 1996; Young et al., 2011). One major difference between these calculations and those of Gaidos et al. (2009) is the use of revised AGB star yields from Karakas (2010). Larger O-isotopic variations could occur in individual SFRs due to the stochastic effects of small numbers of SNe in smaller molecular clouds, however.

Production rates $p_i$ and $q_i$ of secondary isotopes such as $^{17}$O and $^{18}$O increase with metallicity, so $x_i$, $y_i$, and the offset between them will also increase with metallicity. If the yields of the two isotopes depend on metallicity in the same way, the trajectories of the ISM and SFRs in a log-δ three-isotope plot will be parallel and have a slope of one (pure secondary/secondary behavior). In Fig. 3 the slopes deviate from unity, especially at low metallicity. This is because the metallicity-dependence of the production of one heavy isotope differs slightly from the other. Although the ideal nucleosynthetic production of a secondary isotope is proportional to the abundance of the primary isotope, stars of different masses produce the isotopes in different relative abundances and differences can arise that depend on metallicity. For example, metallicity partially controls the opacity of stellar atmospheres and hence stellar evolution and mass loss by winds. If $^{17}$O and $^{18}$O are preferentially produced during different phases of stellar evolution, differences in metallicity will produce differences in the ratio of $^{17}$O to $^{18}$O yields. For example, the mass ejected in the $^{17}$O-enriched winds of AGB stars increases with metallicity while no analogous enhancement occurs for $^{18}$O-enriched SN ejecta. Figure 4 shows that the metallicity dependencies of the IMF-integrated yields of $^{17}$O and $^{18}$O are neither identical nor directly proportional to each other, thus at least small departures from slope-one are expected. These by themselves, however, appear insufficient to explain the difference between the Solar System and the present-day ISM (Fig. 3). Moreover, trajectories have slopes smaller than one, at least near solar-metallicity, indicating a gradual increase in $^{18}$O/$^{17}$O with time, opposite in sign to that required to explain the solar-present ISM difference.

Equation 6, summed over isotopes, also predicts the *absolute* steady-state abundance of oxygen in solar-metallicity SFRs, *i.e.,* equal to the total IMF-averaged yield $(p+q)/R$. Our adopted set of yields over-predicts the solar value (Asplund et al., 2009) and clearly GCE is not near steady state, i.e. the inflow of fresh gas exceeds the rate at which mass is locked up in stellar remnants. Moreover, some gas in SFRs is fragmented and heated by SNe and Wolf-Rayet stars into hot, tenuous superbubbles and fountains and may permanently escape the disk. We can incorporate this effect in the model by multiplying the last term in Equation 1 by a retention factor $f<1$. The steady-state solution becomes:

$$x_i = \frac{s}{1+s}\left[\frac{p_i}{R} + \frac{f(p_i+q_i)}{R(1+s-f)}\right], \tag{8}$$

$$y_i = \frac{s(p_i+q_i)}{(1+s-f)R}. \tag{9}$$

For $s=0.1$, small values of $f$ will produce a diffuse ISM that is $^{17}$O-enriched with respect to SFRs (Fig. 5). However, the observed offset is between the Sun and current SFRs, the diffuse ISM not being observationally accessible. GCE itself does not depart significantly from a slope-one trajectory in a three-isotope plot and we conclude that a heterogeneous ISM cannot, by itself explain the observed offset in $^{17}$O/$^{18}$O.

**Time-dependent star formation**

If the secondary isotopes are produced in different amounts in stars of different masses, which have different main-sequence lifetimes, variation in the rate of star formation can result in an uneven contribution to the inventory of oxygen isotopes (Gaidos et al., 2009). Early calculations of O-isotope GCE, e.g., Timmes et al. (1995), assumed that all three O isotopes are produced in massive stars based on the nucleosynthesis yields available at the time and predicted O-isotopic evolution that closely follows the slope-one behavior expected from the simplest GCE expectations. However, subsequent refinement of nuclear reaction rates has shown that $^{17}$O is not efficiently made in massive stars (Blackmon et al., 1995; Woosley and Heger, 2007) and must largely come from AGB stars and/or classical novae. Both types of sources evolve on longer timescales than do massive stars, indicating that the production of $^{17}$O is delayed with respect to the $^{18}$O made by massive stars. The $^{17}$O/$^{18}$O ratio would thus be expected to increase with time as longer-lived stars evolve away from the main sequence and eventually add $^{17}$O-enriched ejecta to the ISM, causing deviations from a slope-one trajectory in a 3-isotope plot.

The AGB yields adopted here (Karakas, 2010) indicate that the largest production of $^{17}$O occurs in stars of > 2M$_\odot$, which have lifetimes <2 Gyr (Fig. 6); departure from quasi steady-state will scale with changes in star formation rate or metallicity over such an interval. In the GCE model of Gaidos et al. (2009), a constant gas infall rate combined with the Schmidt-Kennicut formula for star formation rate (Fuchs et al., 2009) gives rise to a decreasing $^{18}$O/$^{17}$O ratio with time, one sufficient to explain the Solar-ISM difference (Fig. 7). The calculations of Kobayashi et al. (2011), which included up-to-date nucleosynthesis yields for both AGB stars and supernovae, also predict an increase (~5%, Figs. 7-8) in $^{17}$O/$^{18}$O over the past 4.5 Gyr, but not enough to explain the Solar-ISM difference (~25%, Fig. 1). Note that neither of these models makes predictions for O-isotopic gradients across the galactic disk which could be compared with the observations in Figs. 1 and 2.

The effect of time-delay can be modeled by decomposing the injection of the *i*th isotope into the ISM into instantaneous and delayed parts proportional to the present and past rates of star formation:

$$p_i = c_i R(t) + d_i(t - \tau) R(t - \tau) \qquad (8)$$

where τ is the time delay between the formation of the stars and when they eject the isotope. The first factor in the time-delay term of the RHS accounts for changes in metallicity; the second accounts for changes in the star formation rate. If we assume that (i) time-delay affects $^{17}$O production but not $^{16}$O or $^{18}$O production, (ii) the mean metallicity of the ISM and SFRs has remained constant over the past 5 Gyr (i.e. creating the "G-dwarf problem"), and (iii) that fractional changes in isotopic composition are <<1, then Eqn. 6 can be linearized to give an expression for the fractional change in the $^{17}$O/$^{18}$O ratio of SFRs since the Sun formed:

$$\Delta \frac{^{17}O}{^{18}O} \approx -4.6 \tau f \frac{d}{dt}\left(\frac{d \ln R}{dt}\right), \qquad (9)$$

where *f* is the fraction of $^{17}$O produced in the delayed component, and the derivatives are in Gyr$^{-1}$. The difference depends on the *curvature* of the logarithmic star formation rate and will be positive (i.e. present ISM enriched in $^{17}$O) only if the star formation rate is decelerating (convex). The reason for this can be understood as follows: a decreasing rate of star formation means that the ISM and SFRs will receive $^{17}$O-rich gas from longer-lived, earlier-forming stars that are more numerous than the shorter-lived, $^{18}$O-contributed stars that formed more recently. However, to produce a positive difference in the ISM or SFRs between the present and past (i.e., 4.6 Gyr ago), that decrease must be larger now, i.e., star formation is decelerating.

The magnitude of the excursion depends on the rate of deceleration but even if most $^{17}$O production is delayed ($f \sim 1$), Eqn 9 shows that the star formation rate must change on a timescale of no more than a few Gyr to explain the ~25% ISM-solar difference; a constant or exponentially decreasing star formation rate will not affect the $^{17}$O/$^{18}$O ratio. The nominal model of Gaidos et al. (2009) assumes a constant rate of gas infall and the Schmidt-Kennicut law and produces a convex star formation history. The estimated rate of star formation approaches an asymptotic value in the last few Gyr. Star formation in the Kobayashi et al. (2011) is also convex, but gas infall peaks 8.5 Gyr ago, and star formation does not crest until about 2 Gyr ago (K11 Fig. 11). These star formation histories could partially explain the origin of the $^{17}$O/$^{18}$O offset predicted by the models.

Similar effects may also occur if the metallicity of the ISM has substantially changed in the past 5 Gyr. If the yield of $^{17}$O, as a secondary isotope, scales with metallicity then the change in $^{17}$O/$^{18}$O due to metallicity evolution and the time delay of $^{17}$O production is approximately:

$$\Delta \frac{^{17}\text{O}}{^{18}\text{O}} \approx -10.6 \tau f \frac{d}{dt}\left(\frac{d[\text{Fe/H}]}{dt}\right)$$

Again, positive evolution in $^{17}$O/$^{18}$O requires decelerating change (in metallicity), which is plausible if GCE if the age-metallicity relationship is flat at recent time (the so-called "G dwarf problem"). Even if $f \sim 1$, a deceleration of ~0.07 dex/Gyr is required over the past 4.6 Gyr to explain the observed offset. Thus, both decelerations in star formation rate and metallicity evolution may have contributed to the isotopic offset between the Sun and the ISM.

Sufficiently rapid changes in star formation rate and metallicity could occur during episodes of enhanced star formation ("starbursts"), specifically around the time of the Sun's formation. Clayton (2003) invoked the merger of a metal-poor satellite galaxy with the Milky

Way 5-6 Gyr ago and a subsequent burst of star formation to explain the statistics of silicon isotopes in presolar SiC stardust grains. Such an event would also leave its mark on oxygen isotopes (Clayton, 2004); elevated levels of star formation immediately prior to the Sun's formation could enrich the galactic disk with $^{18}$O before the subsequent post-main sequence evolution of AGB stars would return the isotopic trajectory to the steady-state situation (Gaidos et al., 2009 Fig. 2).

Figure 7 plots an illustrative but unrealistically simple case where a single generation of stars forms instantaneously from a gas of nearly solar metallicity and 4 Myr later begins adding oxygen back to the gas. No star formation occurs subsequent to the burst. The IMA is used and we do not differentiate between the diffuse ISM and SFR phases. The yields are the same as in the previous calculations. To map GCE onto a three oxygen-isotope plot it is also necessary to specify the fractional mass processed by this generation of stars and the epoch at which the Sun forms relative to the starburst. Values of 3% and 9 Myr, respectively, cause the oxygen isotopic composition of the ISM to evolve to its current $^{17}$O-rich position 4.6 Gyr after the formation of the Sun. In the first 10 Myr after the starburst, the isotopic trajectory of the ISM is towards $^{16}$O-rich conditions as the most massive stars explode. During the subsequent 15 Myr the ISM is enriched with $^{17}$O and $^{18}$O by SN of 12-18 $M_\odot$ progenitors. At around 25 Myr, "super" AGB stars begin evolving off the main sequence and ejecting $^{17}$O-enriched gas. This phase continues as "normal" AGB stars continue to add $^{17}$O-rich material and is largely complete by 2 Gyr after the starburst. Obviously, the IMA should be relaxed and a burst must be superposed on a background level of star formation and metal-poor gas infall. However, we find that these trajectories are sensitive to yields that are very uncertain and from phases of progenitors which have not yet been established. Specifically, nucleosynthetic models predict that the yields of very massive (>40 $M_\odot$) stars are sensitive to uncertain parameters including the degree of fallback onto the remnant black hole or neutron star, the effects of mass loss, and the effects of rotation (Meynet et al., 2010; Woosley and Weaver, 1995) and the existence of "super" AGB phases (7-12 $M_\odot$) has yet to be established. Our calculations show that the $^{17}$O-richness of the ISM *could* be explained by the delayed contribution of AGB stars to the ISM, but a quantitative test of any model must await a sounder foundation of yield calculations.

**Contribution of classical novae to O isotope GCE**

Classical novae may also be important but poorly quantified contributors to "delayed" $^{17}$O. Novae are thermonuclear explosions occurring on white dwarfs (WDs) due to accretion of H and He from less-evolved binary companions. Overall, they eject relatively little mass into the ISM and hence are unimportant for the GCE of most isotopes. However, the high-temperature H-burning that occurs is predicted to produce very large amounts of the light isotopes $^7$Li, $^{13}$C, $^{15}$N and, of importance here, $^{17}$O, as well as the radioactive nuclei $^{26}$Al and $^{22}$Ne (José and Hernanz, 1998; Starrfield et al., 1998), and thus may play an important role in the galactic production of all of these (Romano and Matteucci, 2003). Because they require both the evolution of the low- or intermediate-mass parent star to the WD stage and ~1-2 Gyr of WD cooling to ensure a strong nova outburst (Romano and Matteucci, 2003; Romano et al., 1999), novae are expected to first make a significant contribution to GCE several Gyr after AGB stars. This may lead to a delayed elevation of the galactic $^{17}$O/$^{18}$O ratio, one that might help explain the difference between the Solar composition and that of the present ISM.

Novae are rarely included in models of GCE due both to their unimportance in the production of most elements and to the poor understanding of many key parameters, including the fraction of WDs in nova systems throughout galactic history, the total mass ejected by novae and, critically, the nucleosynthetic yields. To our knowledge, there is only one study addressing the role of novae in the GCE of O isotopes (Romano and Matteucci, 2003). These authors incorporated novae into a standard (i.e. single-phase ISM) numerical GCE model and used the observed present-day nova rate in the Galaxy to constrain a parameterization of the nova rate in the past. They considered a range of nucleosynthetic prescriptions for low-mass stars, supernovae and novae. Unfortunately, $^{18}$O was not included in this model, so we cannot directly infer predictions for the GCE of the $^{17}$O/$^{18}$O ratio. However, since novae are not thought to be a significant source of $^{18}$O (José et al., 2012), we can gain some insight by comparing the predicted $^{17}$O/$^{16}$O ratio evolution of Romano and Matteucci (2003), including novae, to other models of $^{18}$O/$^{16}$O evolution calculated with GCE models not including novae.

Figure 8 compares the time evolution of the O isotopic ratios in the solar neighborhood predicted by several GCE models. The isotopic trends have all been "renormalized" so that they are forced to have solar composition at the time of solar birth. As discussed at length by Timmes and Clayton (1996), the renormalization procedure compensates for the large uncertainties in predicted absolute isotopic abundances, and allows one to compare relative isotopic trends with each other and with high-precision data. On this plot, canonical slope-one evolution corresponds to identical trends for $^{17}O/^{16}O$ and $^{18}O/^{16}O$, as observed in Fig. 8 for the model of Timmes et al. (1995), which was based on pure secondary synthesis of $^{17}O$ and $^{18}O$ in massive stars and for which the curves are indistinguishable on the plot.

Two predictions of Romano and Matteucci (2003) for the evolution of $^{17}O/^{16}O$ are shown in Fig. 8. Both models assume that $^{17}O$ is produced solely by novae, so these calculations provide a limiting case since some $^{17}O$ is certainly made by AGB stars. The lower trend ("Nova-2") assumes that $^{17}O$ synthesis in novae is primary (all novae have the same $^{17}O$ yield, regardless of the metallicity of the progenitor stars) whereas for the upper one ("Nova-1") the nova yields are scaled according to metallicity. Both models predict a stronger increase in $^{17}O/^{16}O$ since solar birth than GCE models that do not include novae. The effect of novae on the evolution of the $^{17}O/^{18}O$ ratio can be estimated by comparing these trends to the plotted $^{18}O/^{16}O$ trends. Taking the Timmes et al. (1995) calculation as representative of $^{18}O/^{16}O$ evolution and the "Nova-1" GCE trend for $^{17}O/^{16}O$ gives the most extreme increase in $^{17}O/^{18}O$ since solar birth: ~70%. In contrast, considering the "Nova-2" $^{17}O/^{16}O$ trend and the Kobayashi et al. (2011) $^{18}O/^{16}O$ trend would suggest a much smaller increase of ~10%. Thus, the limited modeling work done to date suggests that nova production of $^{17}O$ over the past 4.6 Gyr of GCE may readily explain the observed difference in $^{17}O/^{18}O$ between the Sun and local ISM, but this depends on many uncertain details and it will require more investigation. Moreover, there is no quantitative model prediction for the shape of a $^{17}O/^{18}O$ gradient across the galaxy expected if novae are primary $^{17}O$ producers, and therefore the molecular cloud observations in Fig. 1 are not yet a good diagnostic test.

## Presolar Grains and O Isotope GCE

Presolar circumstellar grains make up tens to hundreds of ppm of the most primitive meteorites and cometary dust (Nittler, 2003; Zinner, 2007). These micron-sized and smaller grains formed in outflows of evolved stars and were part of the material from which the Solar System formed. They are recognized as pristine stardust by extreme isotopic anomalies, largely reflecting the nuclear processes which occurred in their parent stars. Presolar oxide and silicate grains show huge anomalies in O-isotopic ratios (Figs. 9, 10), with the vast majority of grains being enriched in $^{17}$O and depleted in $^{18}$O, relative to the Solar System. This composition is in qualitative agreement with compositions measured spectroscopically (albeit with large errors) in red giant and O-rich AGB stars and reflects mixing into the stellar atmosphere of material which has experienced H-burning nucleosynthesis through the CNO cycles. The high dust production of AGB stars as well as good agreement between the grain compositions and both observations and predictions of AGB star compositions has been taken as strong evidence that a majority of the presolar O-rich grains formed around low-mass AGB stars (Huss et al., 1994; Nittler et al., 1994; Nittler et al., 2008).

Nittler (2009) performed Monte Carlo modeling of the expected mass and metallicity distributions of parent AGB stars of presolar grains and found that with reasonable parameters, the predicted O-isotopic distribution compared well with the observed distribution. A key result was that achieving good agreement between the model and data required that the initial compositions of the parent stars had solar-like $^{17}$O/$^{18}$O ratios. This reflects the fact that the "dredge-up" process that mixes the ashes of H-burning into a red giant envelope always increases the $^{17}$O/$^{18}$O ratio, since CNO-cycle nucleosynthesis enhances $^{17}$O and destroys $^{18}$O. Monte Carlo models assuming higher initial $^{17}$O/$^{18}$O ratios, like those observed in the present-day ISM, provided poor matches to the data, particularly because a fraction of the grain population has $^{17}$O/$^{18}$O between solar and the current ISM.

We suggested above that GCE including the role of novae in producing Galactic $^{17}$O might explain the O-isotopic difference between the Sun and local ISM. If so, this would imply a continuous increase in $^{17}$O/$^{18}$O over the history of the Galaxy, whereas Nittler (2009) assumed a constant (solar) $^{17}$O/$^{18}$O ratio with time. Fig. 9 shows the results of a Monte Carlo model in which we simulate nova-influenced GCE by assuming that the $^{17}$O/$^{18}$O ratio increases linearly with metallicity and has the solar value at the time of solar birth. Other parameters are very similar to those which provided a good match to the presolar grain data in the original work of Nittler (2009), namely an average age-metallicity relationship is assumed to exist in the Galaxy, the Sun is assumed to have either a slightly unusual metallicity for its age or equivalently to be slightly $^{16}$O-rich for its metallicity, and there is assumed to be a 1-σ spread of ~12% in metallicity for stars born at the same time. This spread in metallicity and concomitant spread in O-isotopic ratios is at least qualitatively consistent with the observed metallicity and isotopic spread in stars and MCs in the solar neighborhood. As shown in Fig. 9, this model is in good agreement with the observed distribution of O isotopic ratios in the grains and provides further support to the contention that the presolar galactic ISM had $^{17}O/^{18}O$ ratios significantly lower than does the present-day ISM. Note that this model does not rule out local SN enrichment of the protosolar cloud. In fact, the unusual metallicity and/or $^{16}O$ contents of the Sun required for a good fit (Nittler, 2009) may well point to enrichment by relatively massive SNe which would not greatly shift the $^{17}O/^{18}O$ ratio.

The presolar grain data may provide additional insights into the question of whether the solar O-isotopic composition was strongly influenced by a local enrichment by SN ejecta relatively soon before formation. Astronomical estimates of the timescales of circumstellar dust injection into the ISM and interstellar dust destruction, as well as the overall amount of dust in the Galaxy have led to the inference that most interstellar dust is formed in the ISM itself, probably by low-temperature accretion and coagulation in molecular clouds (Draine, 2009; Dwek, 1998; Hirashita, 2010; Zhukovska et al., 2008). As a first approximation, such grains should have the bulk O-isotopic composition of the ISM at the time of their formation. Zhukovska et al. (2008) estimate that isotopically normal ISM silicates should have been at least 100 times more abundant than AGB silicates at the time of solar birth. Dust destruction timescales are of order $10^8$ yr (Jones et al., 1996), much longer than the lifetime of star-forming regions. Hence, if the solar $^{17}O/^{18}O$ ratio composition was indeed strongly modified by a very late injection of supernova material (Young et al., 2011), one would expect to find a significant population of grains in primitive meteorites with the typical isotopic composition of the pre-injection protosolar ISM. Such a population would dominate the distribution of isotopically anomalous grains identified as presolar, and have $^{17}O/^{18}O$ ~0.25 and a narrow range of $^{17,18}O/^{16}O$ ratios.

Fig. 10A shows $^{17}O/^{18}O$ ratios plotted against $^{18}O/^{16}O$ ratios for ~1400 presolar O-rich grains; probability contours are overlain on the data in Fig. 10B. There is a single, very broad maximum in the probability density centered on $^{18}O/^{16}O=1.9\times10^{-3}$ ($^{16}O/^{18}O=526$) and $^{17}O/^{18}O=0.28$ ($^{18}O/^{17}O=3.6$). It is unlikely that this represents a population of ISM-formed grains with a pre-SN-enrichment O-isotopic composition for a number of reasons. First, as discussed above, the distribution of O-isotopic ratios in the grains is completely consistent with expectations based on stellar evolution and GCE theory developed independently of the grains (Nittler, 2009) so the maximum in the probability distribution is not an extraneous feature in demand of explanation. Second, only ~4-5% of the grains lie within the contour corresponding to 90% of the maximum probability, compared to the expectation that interstellar dust should outnumber circumstellar dust by a factor >100 (Zhukovska et al., 2008). Third, the peak occurs at a location of the O three-isotope plot where two distinct evolutionary trends (so-called Groups 1 and 2, Nittler et al., 1997) are coming together and thus a pile-up of grains would naturally be expected to occur. Fourth, this region of the plot includes refractory oxides like $Al_2O_3$, and $MgAl_2O_4$. Since grain growth in the ISM is believed to occur by a non-thermal low-temperature accretion process (Draine, 2009), such high-temperature phases are extremely unlikely to form in the ISM and an origin as circumstellar condensates is strongly favored. Fifth, the probability peak occurs at a more $^{17}$O–rich composition than the preferred current local ISM composition (Wouterloot et al., 2008; Young et al., 2011). Therefore, the distribution of O-isotopic ratios in presolar oxide and silicate grains provides an additional argument that the solar $^{18}$O/$^{17}$O ratio is not drastically different from typical values in the Galactic ISM at the time of solar birth. A potential experimental test would be measurements of isotopic compositions of additional elements besides oxygen in grains with compositions close to the maximum in the probability distribution. If such grains showed strong nuclear anomalies in other elements, this would be evidence that they were indeed circumstellar, not interstellar, grains.

## GCE and Solar System Oxygen Heterogeneity

Three decades ago D. D. Clayton introduced the concept of "cosmic chemical memory" to describe how chemically and isotopically fractionated reservoirs in the interstellar medium could persist through star formation and provide an explanation for isotopic anomalies in meteorites (Clayton, 1982). In particular, Clayton (1988) showed that since isotopic ratios in the Galaxy change with time, interstellar dust components formed at different times would be expected to have isotopic anomalies relative to each other and to the bulk ISM. Chemical and/or physical fractionation of these isotopically distinct components could then potentially influence isotopic variations in solar system materials. For example, he argued that refractory Al-rich dust in the ISM should be older on average than the bulk interstellar gas, and thus relatively $^{16}$O-rich due to GCE. If calcium aluminum inclusions (CAIs) in meteorites preferentially formed from such $^{16}$O- and Al-rich grains, their $^{16}$O-richness, relative to other planetary materials, could be explained.

Although chemical explanations for solar system O-isotopic heterogeneity have been largely favored in recent years, especially self-shielding during UV photodissociation of CO (Clayton, 2002; Lyons and Young, 2005; Yurimoto and Kuramoto, 2004), some recent studies have revisited the "chemical memory" idea. Jacobsen et al. (2007) presented an updated version of the basic Clayton (1988) model, in which they assumed that a fraction of newly synthesized O isotopes are incorporated into dust grains. Because the dust grains are on average older than the gas, and because $^{17,18}$O/$^{16}$O ratios increase with time, an isotopic difference of ~10% between $^{16}$O-rich dust and $^{16}$O-poor gas is predicted to exist at the time of solar system formation. This model hence reproduces the scale of isotopic heterogeneity in the solar system. Figure 11A schematically illustrates the basic concept behind the Clayton (1988) and Jacobsen et al (2007) models. If this mechanism was indeed responsible for the isotopic heterogeneity of the solar system, these models make the prediction that the bulk solar composition (dominated by gas) is $^{16}$O-poor relative to solid material (Young et al., 2008). This is in violent conflict with the recent determination that the solar wind (and presumably the bulk Sun) is enriched in $^{16}$O, relative to planetary materials (McKeegan et al., 2011).

Meyer (2009) suggested a variation of the chemical memory model in which interstellar dust is dominated by recent stellar ejecta (either directly condensed in stellar outflows or accreted onto existing dust grains prior to mixing with average ISM material). In this case (Fig. 11B), the dust would be "young" and $^{16}$O-poor relative to the gas. Krot et al. (2010) presented O-isotopic evidence from igneous CAIs with isotopically fractionated O (so-called FUN and F CAIs) that primordial solids in the solar system may indeed have been $^{16}$O-depleted relative to the solar wind composition. In addition, fine-grained silicates in highly primitive anhydrous interplanetary dust particles (Aléon et al., 2009; Keller and Messenger, 2011; Nakashima et al., 2012) and comet Wild 2 samples (McKeegan et al., 2006), thought to be very primitive samples of nebular dust, have $^{16}$O-poor compositions similar to other planetary materials, also suggesting that primordial dust in the solar system was $^{16}$O-poor, and perhaps supporting the "young dust" GCE chemical memory model.

A difficulty with both the Jacobsen et al. (2007) and Meyer (2009) models is that both assume that fresh circumstellar dust condenses with the bulk O-isotopic composition of a generation of stars and hence follow the average GCE O-isotope trend. However, this assumption is inconsistent with our understanding of the sources of fresh stardust in the Galaxy. Although most newly-synthesized oxygen is injected by supernovae, most of this is gaseous and AGB stars in fact dominate the input of dust to the Milky Way ISM (Kemper et al., 2004). Thus if the average composition of interstellar dust is controlled by young, freshly condensed stardust, it would be expected to be enriched in $^{17}$O, but not $^{18}$O, as observed in and predicted for the vast majority of AGB stars (Karakas, 2010; Smith and Lambert, 1990) and presolar O-rich grains in meteorites (Nittler et al., 1997; Nittler et al., 2008). On the other hand, as discussed in the previous section, astronomical data indicate that most interstellar dust is not fresh stardust, but rather dust that has grown in molecular clouds from homogenized interstellar material (Zhukovska et al., 2008). Therefore interstellar dust is a mixture of both $^{17}$O-rich stardust and isotopically homogenized dust, and the average composition depends on the details and timescales of the formation and destruction of these components.

Fig. 11C illustrates a more realistic chemical memory model based on these considerations. Dust forming in the ISM itself should lie on the overall GCE trend but slightly $^{16}$O-rich relative to the gas as predicted by Clayton (1988) and Jacobsen et al. (2007). In contrast, fresh stardust will be enriched in $^{17}$O and depleted in $^{18}$O, relative to the bulk ISM, so the average interstellar dust should be slightly displaced to $^{17}$O–rich compositions compared to the bulk ISM gas. Of course, quantitative predictions for the compositions of the gas and dust would require much more sophisticated modeling than has been considered so far, taking into account, for example, condensation efficiencies in different types of stars, timescales of grain destruction, re-formation and isotopic equilibration in the ISM, etc. Moreover, for simplicity, the idea outlined here assumes a single-phase ISM, in contrast to our two-phase O-isotope GCE models presented in an earlier section. However, because the lifetimes of interstellar dust grains (Jones et al., 1996) far exceed the lifetimes of star forming regions, we do not think that the basic idea would change, except noting that, because dust growth in the ISM occurs in the dense phases, the ISM-formed dust would be expected to follow the GCE trend of SFRs, not that of the diffuse ISM.

Figure 11C does not obviously support the idea that the Solar System formed from reservoirs of $^{16}$O-rich gas and $^{16}$O-poor dust established by galactic evolution processes, since one would expect the average dust in the protosolar cloud to be $^{18}$O-poor, and possibly $^{17}$O-rich, relative to the bulk gas. However, this analysis considers only average, secular GCE and not the possibility of transients or local enrichment of the Sun's parental material. Figure 11D illustrates one way in which GCE, chemical memory and SN enrichment could combine to make the Sun $^{16}$O-rich, relative to primordial solids in the Solar System. This cartoon assumes that the SFR from which the Sun condensed had an initial gas composition based on GCE and a dust composition slightly $^{17}$O-rich and $^{18}$O-poor, relative to the gas as in Fig 10C. Mixing of supernova ejecta into the cloud shifts the bulk gas composition to $^{16}$O-rich relative to the dust. This scenario differs significantly from previous models (e.g., Prantzos et al., 1996; Young et al., 2011) in that relatively massive supernovae (roughly >25 $M_\odot$) are invoked to shift the composition mostly in the direction of pure $^{16}$O without substantially modifying the $^{17}$O/$^{18}$O ratio.

This supernova enrichment could have occurred in the "starburst" scenario of the Sun's formation proposed by Clayton (2003, 2004) and discussed above. Figure 12 plots the evolution of the ISM as a result of such a starburst with respect to a dust component whose composition lags the gas and is assumed to have a near-solar composition at the time of the Sun's formation. The evolution is initially towards the $^{16}$O-rich corner of a three oxygen isotope plot as massive SNe inject material that is poor in the secondary isotopes. The trajectory does not exactly follow a slope-one line nor is the magnitude of the excursion (assuming a starburst that cycles 3% of gas into stars) sufficient to explain the *Genesis* solar wind-based composition of the Sun (McKeegan et al., 2011). Nevertheless the yields of the most massive stars, including Wolf-Rayet winds, are still very uncertain and the size of any burst is ill constrained, so this may yet provide a self-consistent explanation for the O-isotopic distribution of presolar grains, and the $^{16}$O-poor nature of primordial solar system solids (Krot et al., 2010), if the latter inference is borne out by further work.

## Conclusions

Both theory and observation support the contention that O isotopic ratios have evolved in the Galaxy. The $^{17}O/^{16}O$ and $^{18}O/^{16}O$ ratios of the ISM and molecular cloud cores increase with increasing metallicity, reflecting the primary and secondary nucleosynthesis origin of $^{16}O$ and $^{17,18}O$, respectively. However, the traditional view that the two ratios vary quantitatively in lockstep (*i.e.,* moving along a line of slope one in an oxygen three-isotope plot) is neither unequivocally supported by the data nor a foregone prediction of models of galactic chemical evolution. The two secondary isotopes are produced in different abundances in progenitors of different masses and lifetimes, and so the relative numbers of those stars, as well as the elapsed time since star formation, also matter. The most important production site(s) for $^{17}O$ is not even known with great confidence, though it is most likely in low- or intermediate-mass AGB stars and/or classical novae. Since both of these types of stars evolve on longer timescales than the more massive SN progenitors that produce much of the $^{18}O$, this implies a time delay in the production of $^{17}O$ relative to that of $^{18}O$. If the rate of star formation in the Galaxy has changed over time, then the $^{17}O/^{18}O$ ratio has evolved as well. Unfortunately, models are difficult to test due to large uncertainties in both many of the basic model parameters (e.g. nucleosynthetic yields) and in observational data (e.g., isotopic ratios of molecular clouds derived from double-isotope-ratio isotopomer measurements). Nonetheless, we can shed some light on the questions posed in the Introduction:

*Do current observations support deviations from a universal slope-one (constant $^{18}O/^{17}O$ ratio) GCE?*

A possible gradient in $^{17}O/^{18}O$ ratio of molecular clouds across the Galaxy supports the secular evolution of this ratio. A striking feature of the O-isotopic ratios reported for molecular clouds throughout the Galaxy is their very wide scatter, probably due both to intrinsic variability and substantial analytical uncertainties. This observed scatter further complicates attempts to quantitatively test GCE models. Moreover, it is not possible to equate a spatial (i.e., galactocentric) gradient with a unique trend with time, just as all stars and gas do not fall along a single curve of metallicity vs. time. Therefore, it appears that current astronomical observations and GCE models allow for, but do not require, a GCE explanation for the present-day solar-ISM O-isotope discrepancy. Perhaps the strongest evidence that the solar $^{17}O/^{18}O$ ratio was typical of stars forming at the same epoch (and thus favoring non-slope-one GCE) comes from presolar stardust grains in meteorites. Modeling indicates that the observed O-isotopic distribution of presolar oxide and silicate grains is much more consistent with a consistent increase of the galactic $^{17}O/^{18}O$ ratio than with a ratio that is constant in time and equal to the present ISM value. Moreover, current understanding of interstellar dust evolution suggests that, if the Solar System $^{17}O/^{18}O$ ratio was significantly changed by a late injection of SN ejecta, a large population of isotopically anomalous grains with the isotopic composition of the pre-injection ISM would be observed in meteorites. Such a population has not been seen in extensive searches for presolar grains.

*Can the lower $^{17}O/^{18}O$ ratio of the Sun compared to the present-day ISM be explained by GCE or does it require SN enrichment of the protosolar cloud?*

The simplest quasi-steady-state model of a homogeneous (or heterogeneous) GCE in a two-phase ISM fails to produce the ~25% difference observed between the solar $^{17}O/^{18}O$ ratio and that of present-day molecular clouds and young stellar objects. However, we find that dynamic models with time-dependent star formation and that account for the delay in the production of $^{17}O$ relative to $^{18}O$ can predict such an increase in the galactic $^{17}O/^{18}O$ ratio since the Solar System formed. This is consistent with previous work and indicates that the non-slope-one behavior reported in Gaidos et al. (2009) was due mostly to the time-dependence of star formation rather than the heterogeneity of the ISM. (Gaidos et al. (2009) also used AGB yields which are now deprecated). A particularly dramatic manner in which slope-one evolution can occur is a "starburst" or pulse of star formation occurring close in time to solar formation and the subsequent oxygen isotopic evolution of the ISM has been heavily influenced by AGB stars. We show that non-slope-one GCE depends on the *second-derivative* of the star-formation rate, something that is obviously poorly constrained, as well as very uncertain nucleosynthetic yields, especially from classical novae, "super"-AGB stars and very massive stars. The yields of the last depend on stellar rotation, mass-loss phases in winds, and SN ejecta fallback, none of which well-understood.

*Is there a 'Galactic' explanation for the different O-isotopic compositions of the Sun and planets?*

Photochemistry of carbon monoxide (CO) and self-shielding of $C^{16}O$ is currently a popular explanation for the 5-10% O-isotopic difference between the bulk Sun and most planetary materials. However, there is evidence for the existence of $^{16}O$-poor solids very early in the history of the solar system (Krot et al., 2010), raising the question of whether the separate $^{16}O$-poor and $^{16}O$-rich reservoirs were a product of GCE. The basic "cosmic chemical memory" idea of Clayton (1988) leads to the expectation that primordial solids (dust) in the solar system should be $^{16}O$-rich (and perhaps slightly $^{17}O$–rich), relative to the Sun, in conflict with the observations. However, scenarios can be constructed in which the $^{16}O$-rich contribution of very massive stars, for example from a starburst shortly before solar system formation could lead to $^{16}O$-rich solids and a $^{16}O$-poor bulk Sun, independent of photochemistry. Progress on this front will require better determination of the absolute bulk Sun oxygen isotopic composition i.e., whether it lies on or off the slope-one carbonaceous chondrite anhydrous mineral (CCAM) line on the oxygen three-isotope plot, better understanding of dust formation, evolution and destruction processes and timescales in the ISM, better understanding of the distribution of O isotopes in primitive solar system solids, and as above, better understanding of GCE parameters, especially nucleosynthetic yields.

## Acknowledgements

We would like to thank the organizers of the Workshop on Formation of the First Solids in the Solar System on Kauai, Hawaii, for the invitation to discuss the subject of this paper. We thank Amanda Karakas and Chiaki Kobayashi for sharing calculation results. This work was supported in part by NASA grant NNX10AI63G to LRN.

# Figure Captions

**Fig. 1.** Ratio of $^{18}$O to $^{17}$O measured in molecular clouds (Penzias, 1981; Wouterloot et al., 2008) and young stellar objects (Smith et al., 2009) as a function of distance from the Galactic center. The Wouterloot et al. (2008) study included multiple measurements of some MCs (e.g., Orion KL), and for these we combined the data into a single, averaged data point here so as not to bias the overall distribution. Because estimates of the locations of the studied MCs has changed greatly since 1981, Penzias data are only shown for clouds in common with the Wouterloot et al. (2008) study and the galactocentric distances are shifted to the modern values. Grey dashed line indicates schematically a possible gradient observed in the dataset of Wouterloot et al (2008).YSO data are plotted with 2-σ error bars. The reported uncertainties on the individual MC measurements are typically smaller than 10% relative. The 1-σ uncertainty on the Solar System value (5.2±0.2, Young et al. 2011) is smaller than plot symbol.

**Fig. 2.** Ratios of $^{16}$O to $^{18}$O of molecular clouds derived from radio observations of transitions in different molecules plotted against distance from the Galactic center. CO data were derived following Young et al. (2001) by combining line ratios of $^{13}$C$^{16}$O/$^{12}$C$^{18}$O reported by Wouterloot et al. (2008) with the galactic $^{12}$C/$^{13}$C gradient derived by Milam et al. (2005) from CO observations. One-sigma error bars are estimated from the scatter in CO line ratios and the reported uncertainty in the $^{12}$C/$^{13}$C gradient. Formaldehyde data are from the compilation of Wilson and Rood (1994) and OH data are from Polehampton et al. (2005). The thick black line is the gradient derived by Wilson and Rood (1994); the thin solid curve is an exponential fit to the CO J=1-0 data and dashed curve is an exponential fit to CO J=2-1 data.

**Fig. 3.** Predicted quasi-steady-state oxygen isotope ratios in a two-phase model of the ISM for metallicities of $10^{-4}$ (not seen), 0.01, 0.1 and 1Z$_\odot$. Gas in star-forming regions (solid line plus stars) collects ejecta from stars with lifetimes less than 10 Myr; all longer-lived stars contribute to the diffuse ISM (dashed line plus circles). The dotted line is the expected slope-one line of galactic chemical evolution. The inset shows details around the solar-metallicity

calculations in units of ‰. The calculations used Equations 5 and 6, a star formation efficiency of 10%, and IMF-integrated yields of $^{16}$O, $^{17}$O and $^{18}$O based on the IMF of Kroupa (2002), the AGB yields of Karakas (2010), the "super-AGB" yields of Siess (2010), the yields of Kobayashi et al. (2006) with the corrections of Kobayashi et al. (2011) for SN progenitors between 12 and 40 $M_\odot$, and the wind plus ejecta yields of Portinari et al. (1998) for less massive or more massive SN progenitors.

**Fig. 4.** IMF-integrated yields of $^{17}$O and $^{18}$O for metallicities of $10^{-4}$ (lower left), 0.01, 0.1 and 1$Z_\odot$, showing that the two secondary isotopes do not exactly track each other. Solid lines are the contributions of stars more massive than 15 $M_\odot$, which have main-sequence lifetimes less than 10 Myr and thus (hypothetically) contribute to star-forming regions, while dashed lines are for less massive, longer-lived stars that contribute to the diffuse ISM. The standard model (triangles) uses the same sets of yields as in Fig. 3. Lines with crosses use the SN yields of Woosley and Heger (2007) at solar metallicity, and circles use the SN yields of Woosley and Weaver (1995) for progenitors up to 40 $M_\odot$.

**Fig. 5.** Predicted galactic isotopic evolution as in Fig. 3, except only 10% of mass in star-forming regions in returned to the diffuse ISM. Unlike the closed model, this reproduces the total oxygen abundance and the $^{18}$O/$^{16}$O ratio of the Sun (i.e., a star-forming region at solar metallicity). This produces a 30% elevation in the diffuse ISM $^{18}$O/$^{17}$O ratio relative to the Sun. However, the observed offset is between the Sun and molecular clouds and protostellar cores, i.e. star-forming regions; there is too little CO in the diffuse ISM to measure isotopic abundances.

**Fig. 6.** Yields of $^{17}$O (solid) and $^{18}$O (dashed) per unit log interval of time in Gyr vs. main sequence lifetime of the source stars. This plot illustrates that, while most $^{18}$O is injected within 20 Myr of star formation, a large fraction of $^{17}$O is not introduced into the ISM until after ~1 Gyr. The same yields as Fig. 3 were used, except those of Portinari et al. (1998) which were excluded.

**Fig. 7.** Predicted O-isotope ratios from three GCE models, normalized to the values

predicted at solar metallicity. Ellipse shows present-day local ISM based on observations of molecular clouds (estimated from Figs. 1 and 2). Open diamonds are from the model of Gaidos et al. (2009) and circles are from the model of Kobayashi et al. (2011); symbols are plotted at 500 Myr intervals. These models deviate from canonical slope-one evolution (solid line) expected if secondary isotopes $^{17}$O and $^{18}$O are produced at identical rates in the Galaxy. Thick curve is the predicted single-phase GCE in a hypothetical "star burst" scenario in which 3% of the mass of the disk instantaneously forms a generation of stars from nearly-solar metallicity gas around 4.6 Gyr ago. Subsequent isotopic evolution is dictated solely by the post-main sequence evolution of these stars and ejection of O-isotope-enriched gas. The Instantaneous Mixing Assumption (IMA) is adopted in the calculations and the Sun forms 9Myr after the main burst and before AGB stars begin ejecting $^{17}$O-rich material into the ISM. The choice of 3% and 9 Myr is purely to show that that the magnitude of the excursion is sufficient to produce the observed offset between the Sun and the present star-forming regions/ISM, although the scenario itself is unrealistically simplistic.

**Fig. 8.** Predicted evolution of O-isotopic ratios in the solar neighborhood in the last 6 Gyr from three GCE models: TWW95: model of Timmes et al. (1995); $^{17}$O/$^{16}$O and $^{18}$O/$^{16}$O trends lie on top of each other, indicating "slope-1" behavior on an O 3-isotope plot. K11: Model of Kobayashi et al. (2011); $^{17}$O/$^{16}$O evolves slightly more rapidly since solar birth than $^{18}$O/$^{16}$O due to delayed input of $^{17}$O from AGB stars (Fig. 7). Nova-1 and Nova-2: GCE model of Romano and Matteucci (2003) assuming all $^{17}$O production is from novae. Nova-1 is case where nova $^{17}$O production is assumed to be secondary, Nova-2 corresponds to primary nova production of this isotope.

**Fig. 9.** O isotopes measured in presolar oxide grains (Nittler et al., 2008 and references therein) and predicted by a Monte Carlo simulation of the distribution of low-mass AGB stars present at the time of solar birth, following Nittler (2009). O-isotopes are assumed to evolve in the Galaxy along the curve labeled "GCE," in which the $^{18}$O/$^{16}$O ratio increases linearly and the $^{17}$O/$^{16}$O ratio quadratically with metallicity. Labeled ellipses indicate grains believed to have originated in AGB stars undergoing cool-bottom processing (Nollett et al., 2003) or in supernovae and for which their O-isotopes are thus not expected to be explained

by the Monte Carlo model. Note that most O-rich red giants and AGB stars with spectroscopic measurements have $^{17}$O–rich compositions that overlap those of the grains (e.g., Harris and Lambert, 1984; Smith and Lambert, 1990).

**Fig. 10.** Oxygen isotope ratios ($^{17}$O/$^{18}$O and $^{18}$O/$^{16}$O) in presolar oxide and silicate grains (Messenger et al., 2003; Nguyen et al., 2010; Nittler et al., 2008; and references therein). Short-dashed lines indicate solar (terrestrial) values and long-dashed lines indicate $^{17}$O/$^{18}$O of local ISM (Smith et al., 2009; Wouterloot et al., 2008). A) Grain data with 1-σ error bars; box indicate region plotted in panel B. B) Probability contours overlaid on grain data inferred for the grain data by summing Gaussian distributions for each datum with widths taken from analytical error bars. Contour levels range from 10 to 90% of maximum probability level.

**Fig. 11.** Cartoons illustrating various "cosmic chemical memory" scenarios leading to differences in interstellar gas and dust O-isotopic compositions. A) In the model of Clayton (1988) and Jacobsen et al. (2007), $^{17,18}$O/$^{16}$O ratios of both gas and dust increase monotonically with time, but since the interstellar dust is assumed to be older than the gas on average it is slightly $^{16}$O-rich. B) In the "young dust" model of Meyer (2009), interstellar dust is assumed to be dominated by fresh stellar ejecta and is hence $^{16}$O-poor relative to the average gas. C) A more realistic model in which interstellar dust is assumed to be a mixture of $^{17}$O-rich, $^{18}$O-poor circumstellar stardust (mostly from AGB stars) and homogenized dust formed in molecular clouds. D) Same model as in C, but with subsequent mixing of $^{16}$O-rich SN ejecta with the average interstellar gas to produce a solar composition $^{16}$O-rich relative to primordial solids.

**Fig. 12:** Evolution of the oxygen isotopic composition of the gaseous ISM relative to a dust component that lags the gas by 100 Myr (the exact value is unimportant here), following the starburst modeled in Fig. 7. The time elapsed after the starburst is marked at 1 Myr intervals. The effect is to produce a short-lived enrichment in $^{16}$O, followed by a much longer-lived relative depletion before the ISM becomes $^{17}$O-rich from AGB stars (Fig. 7). The composition of the solar wind as sampled by the *Genesis* spacecraft (McKeegan et al., 2011) is indicated; the bulk solar composition is believed to be more $^{17}$O and $^{18}$O-rich than this measured value and lie somewhere on a slope-1/2 line (dashed line), perhaps where it intersects the slope-1 line defined by calcium-aluminum rich inclusions in meteorites (dotted line). The maximum $^{16}$O enrichment produced by this model, about 50 ‰, is close to what is needed to explain the difference between the Sun and the planets, but is highly dependent on model parameters and uncertain supernova yields.

Figure 1

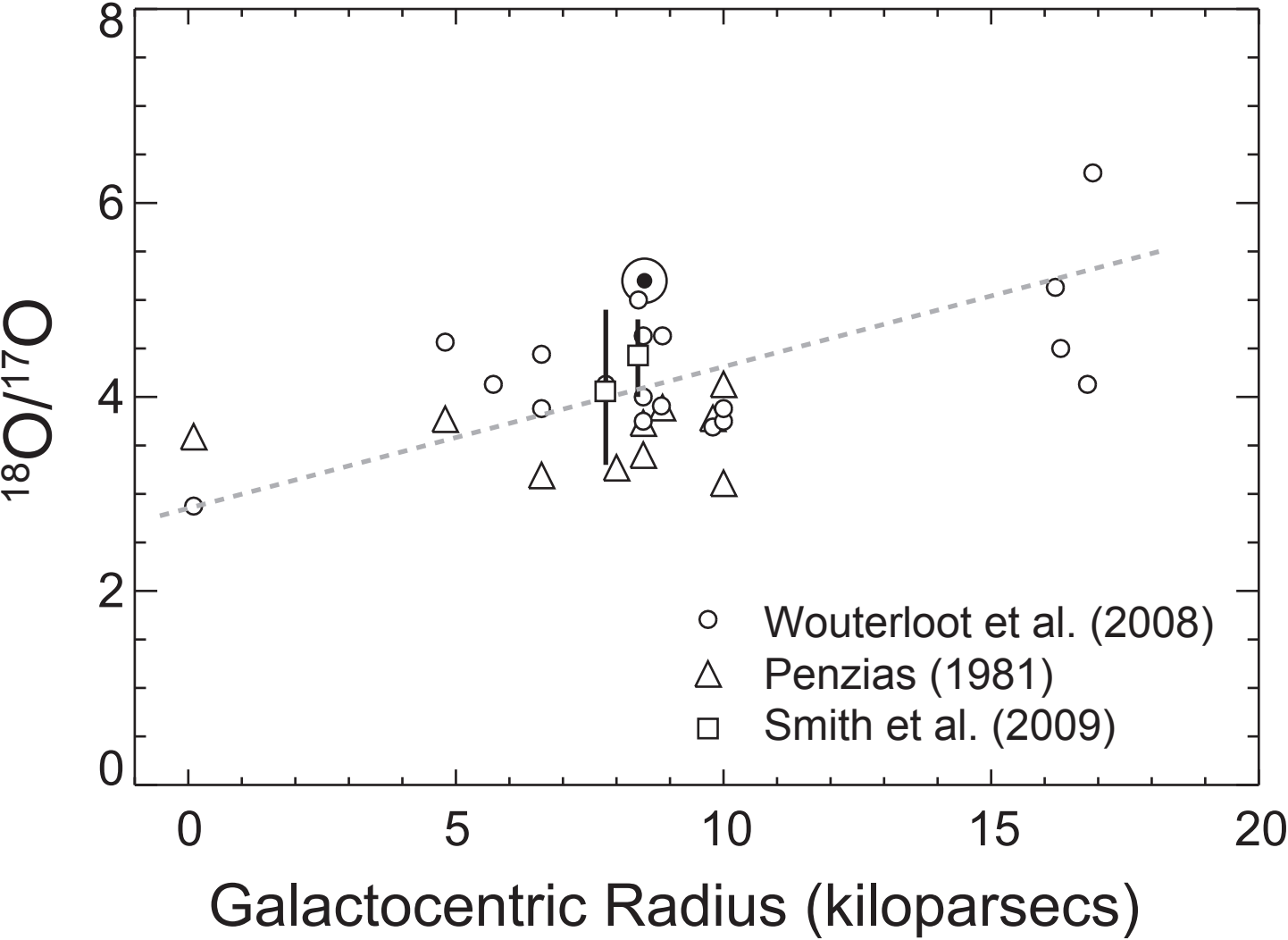

Figure 2

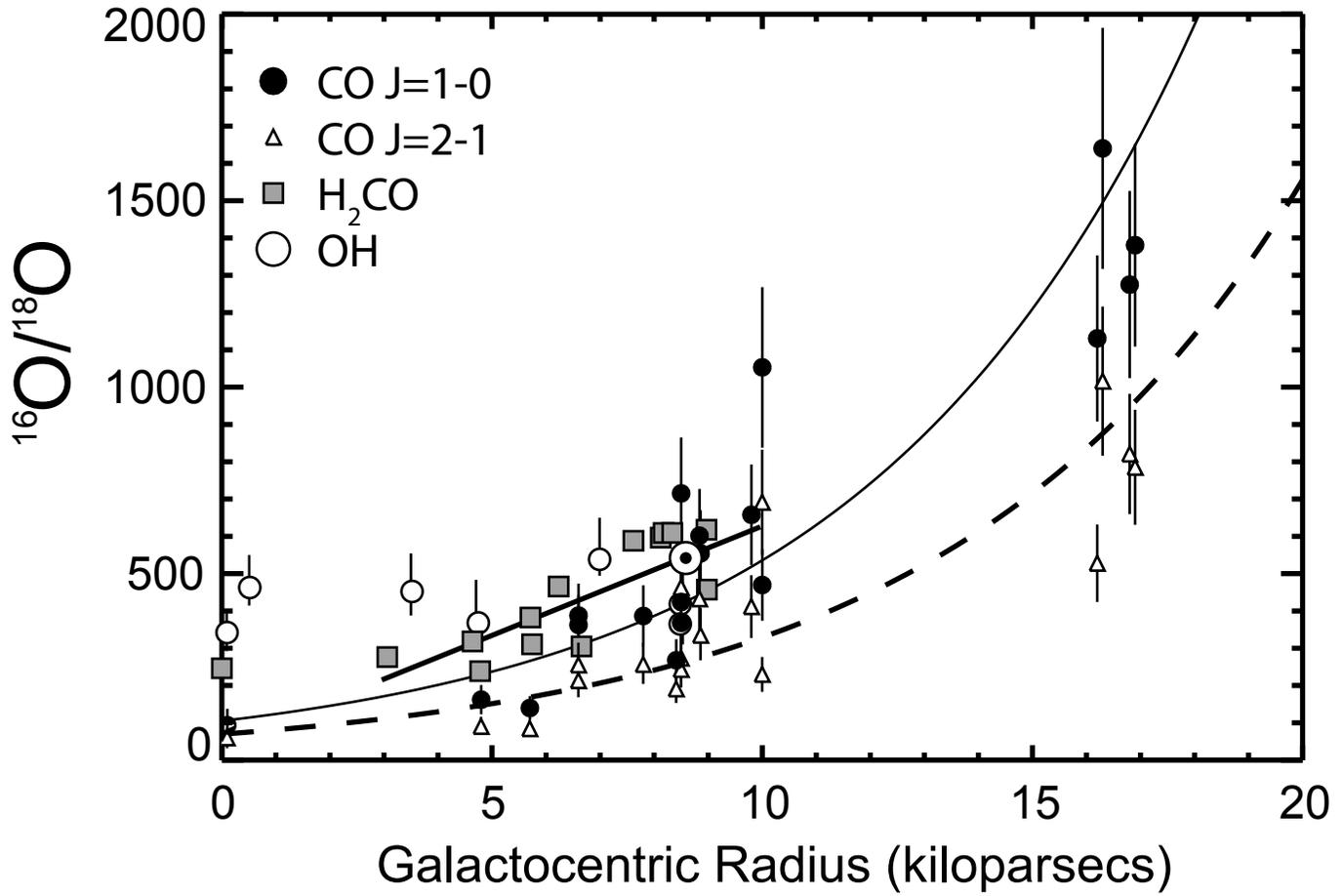

Figure 3

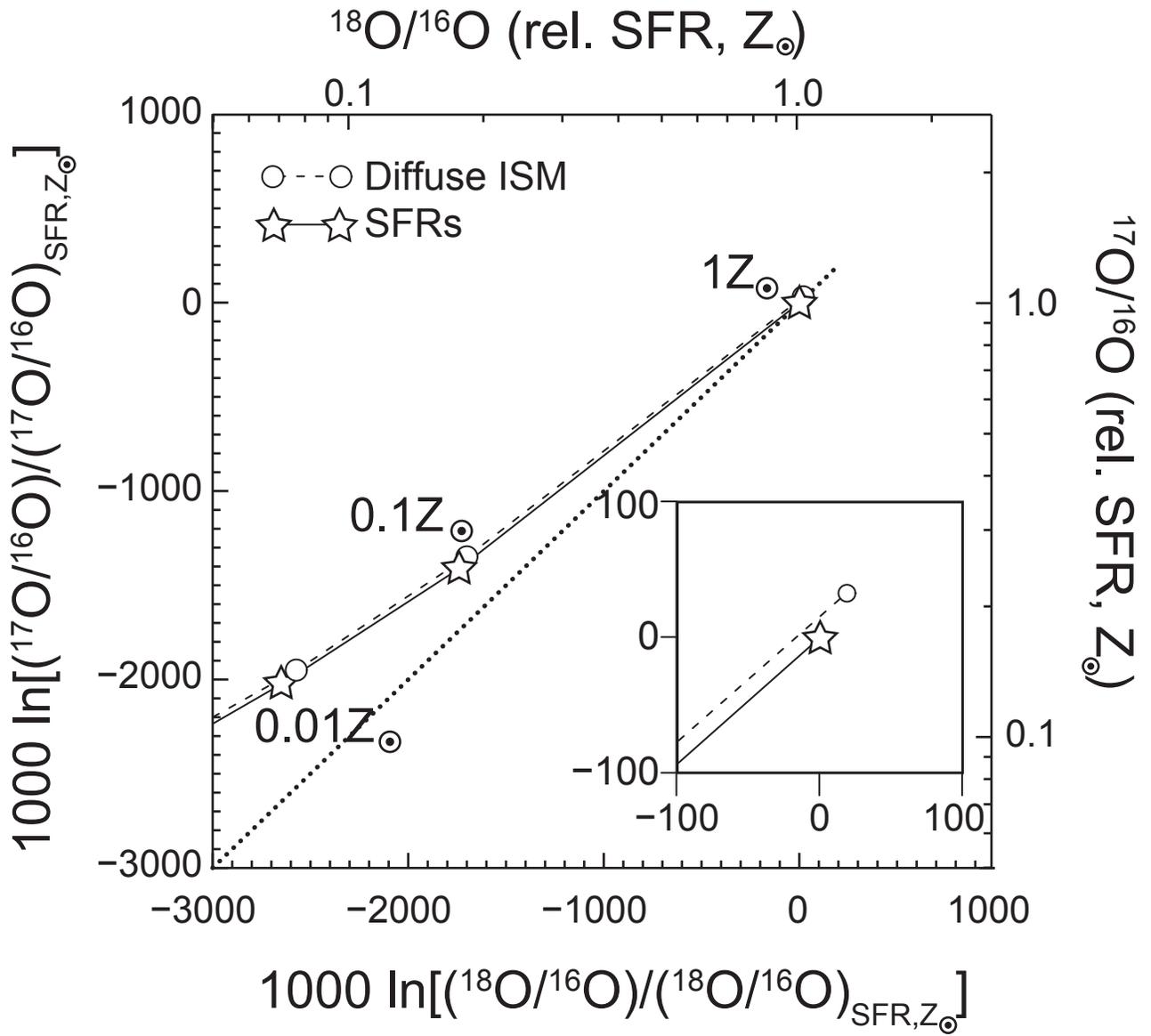

Figure 4

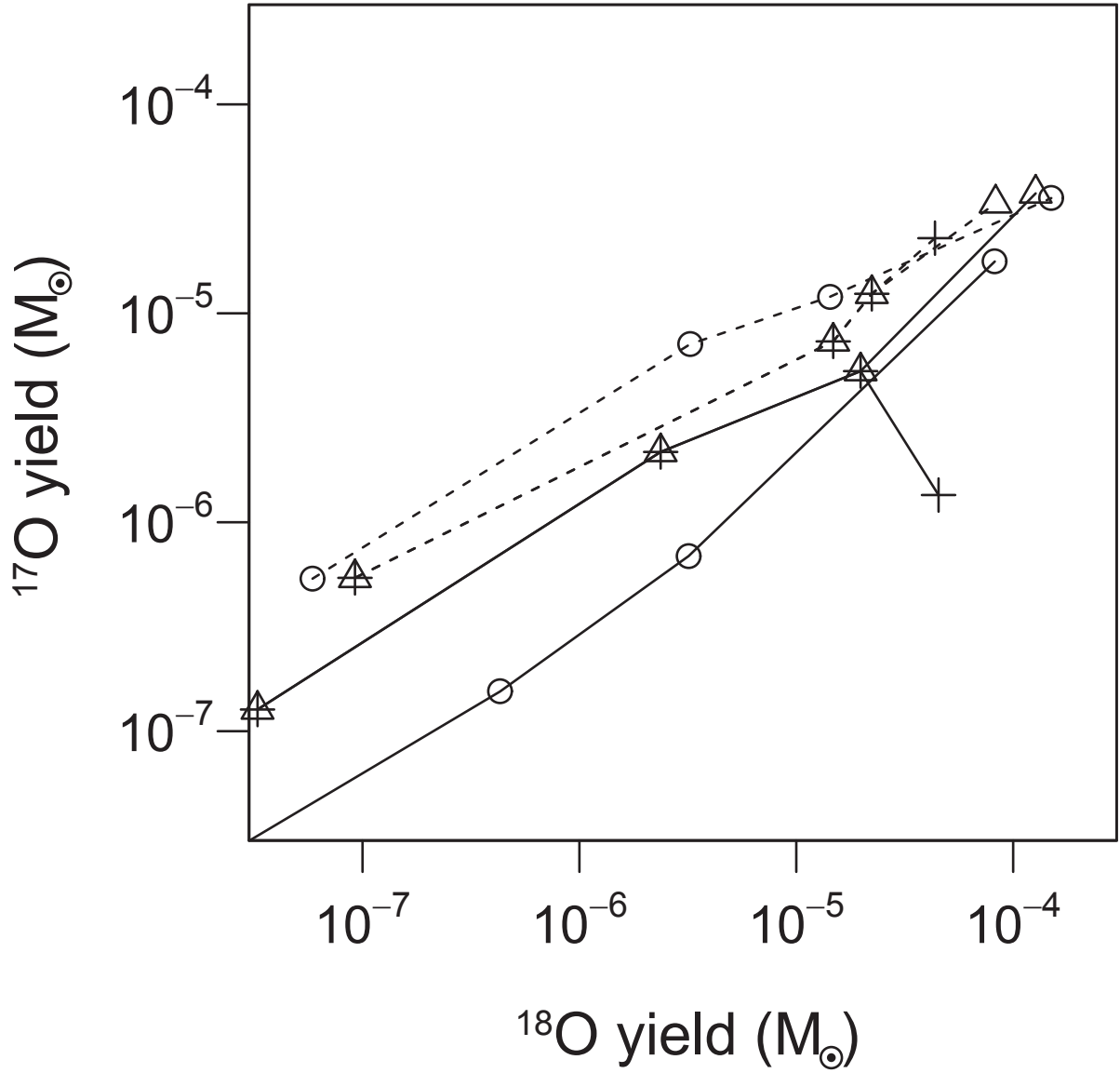

Figure 5

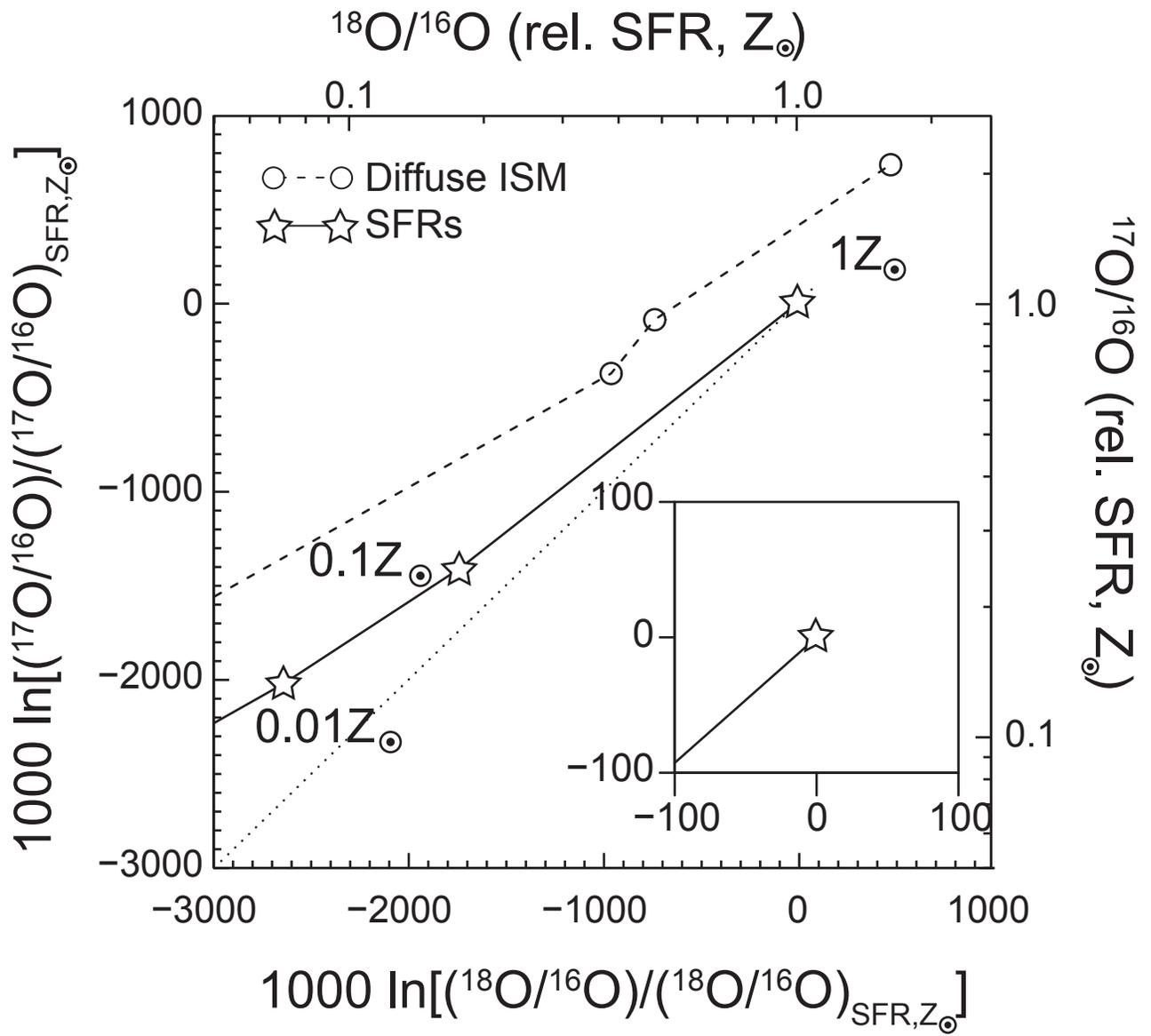

Figure 6

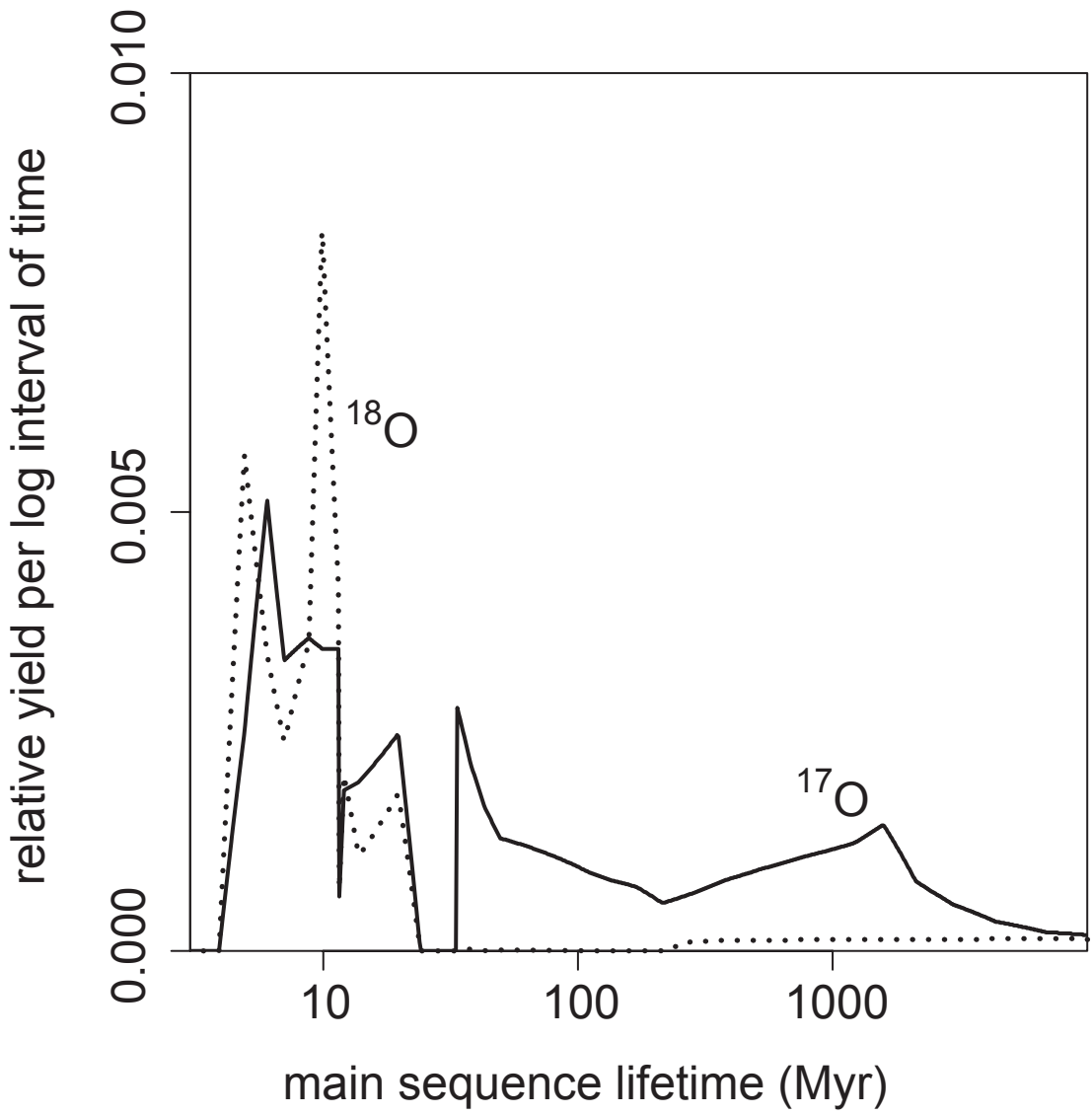

Figure 7

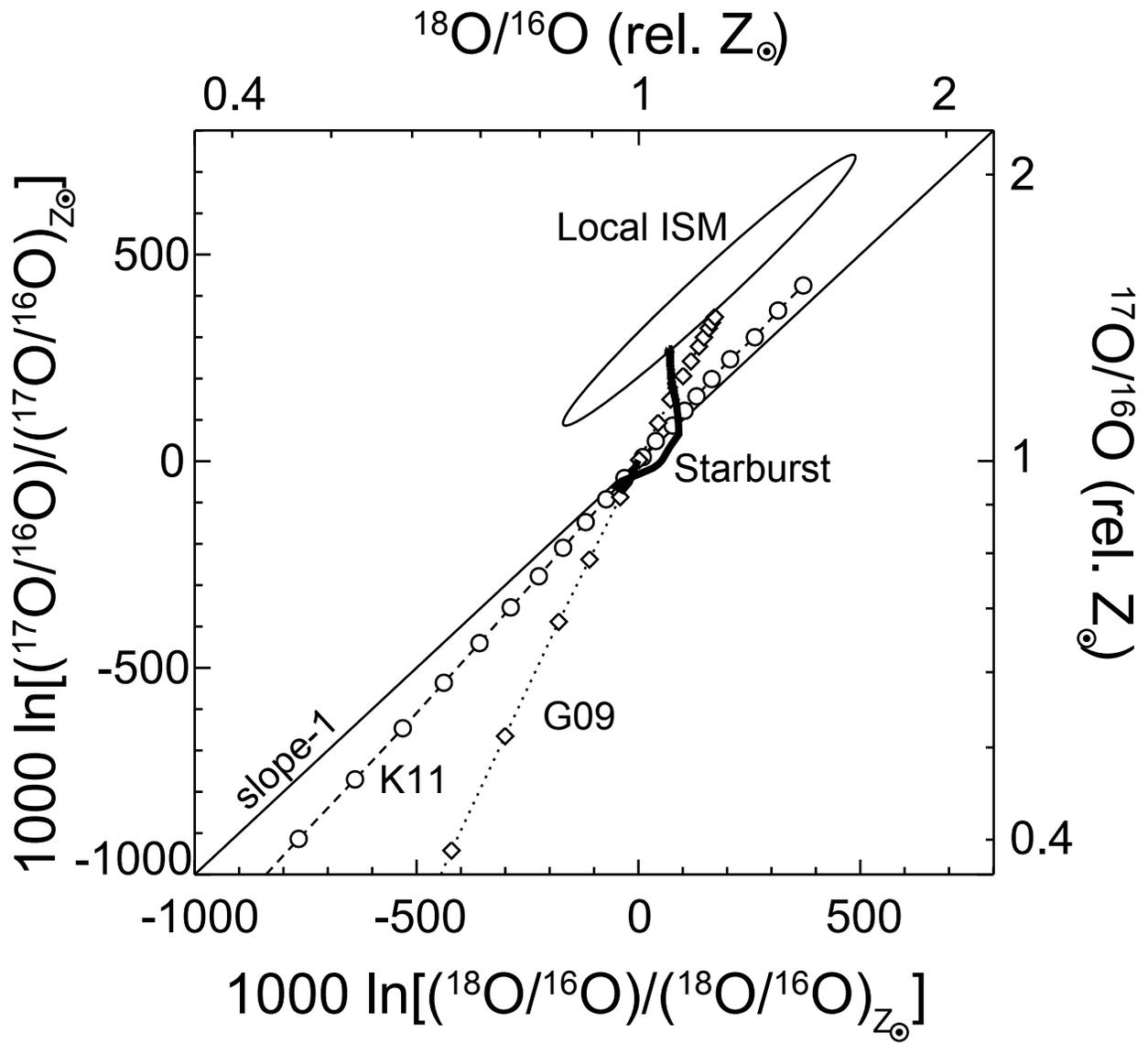

Figure 8

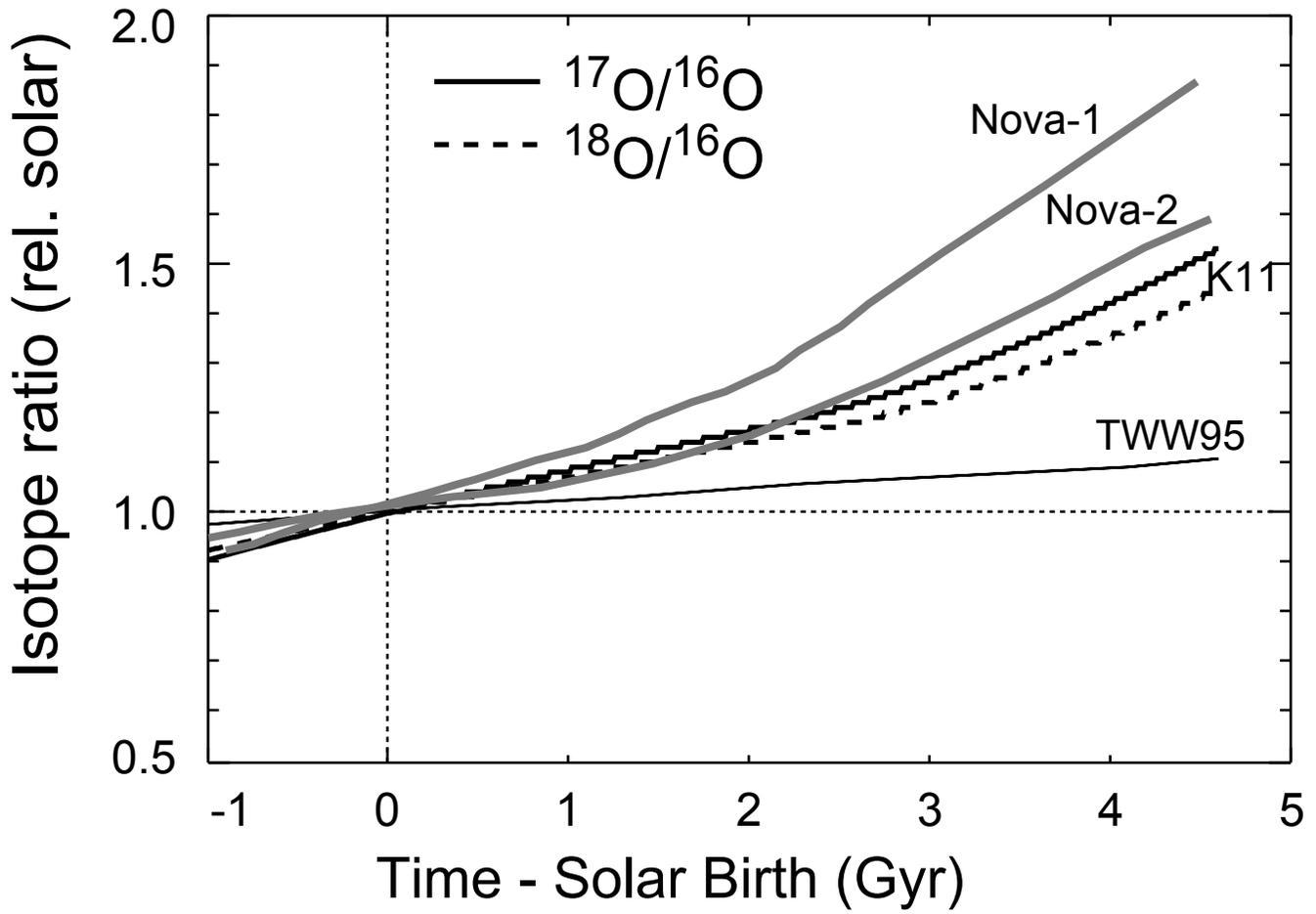

Figure 9

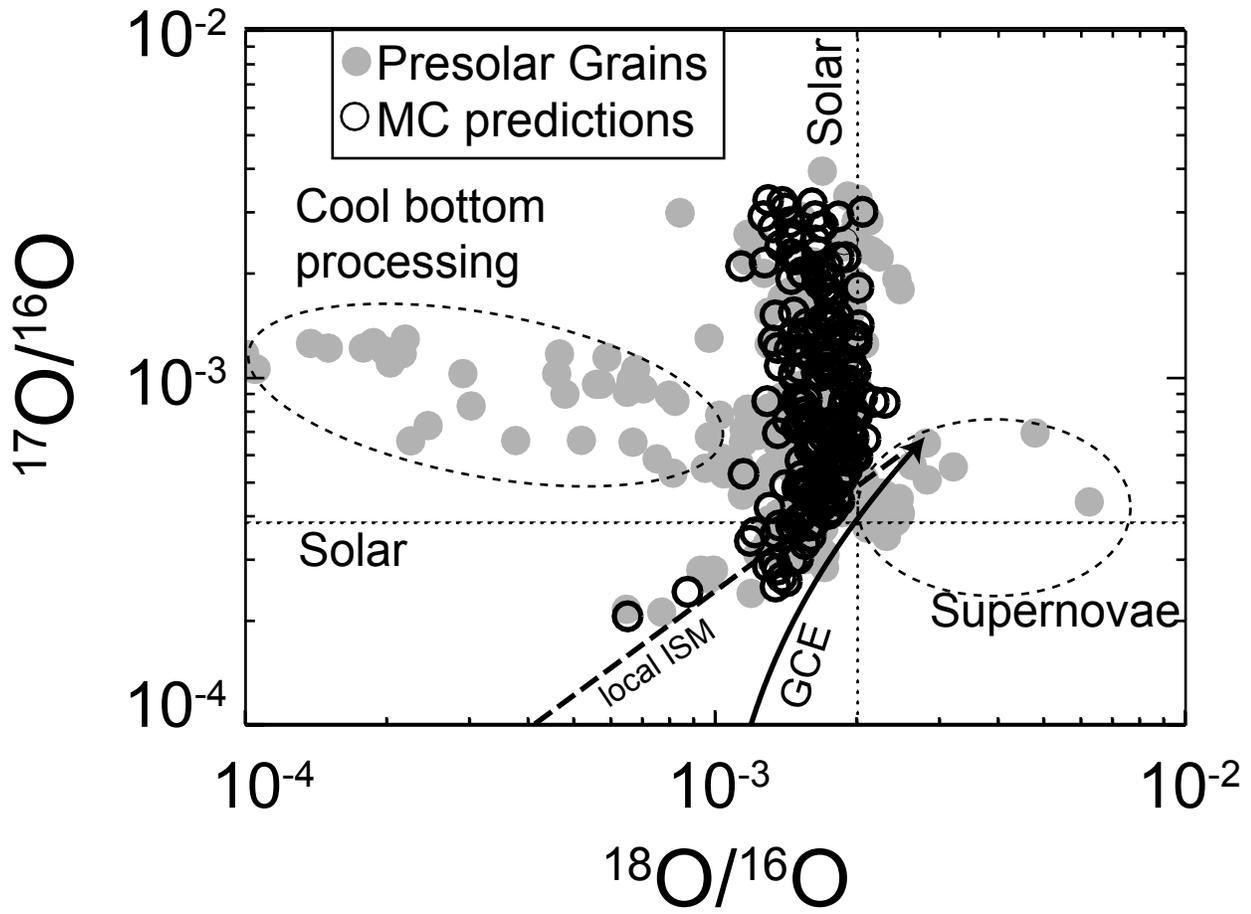

Figure 10

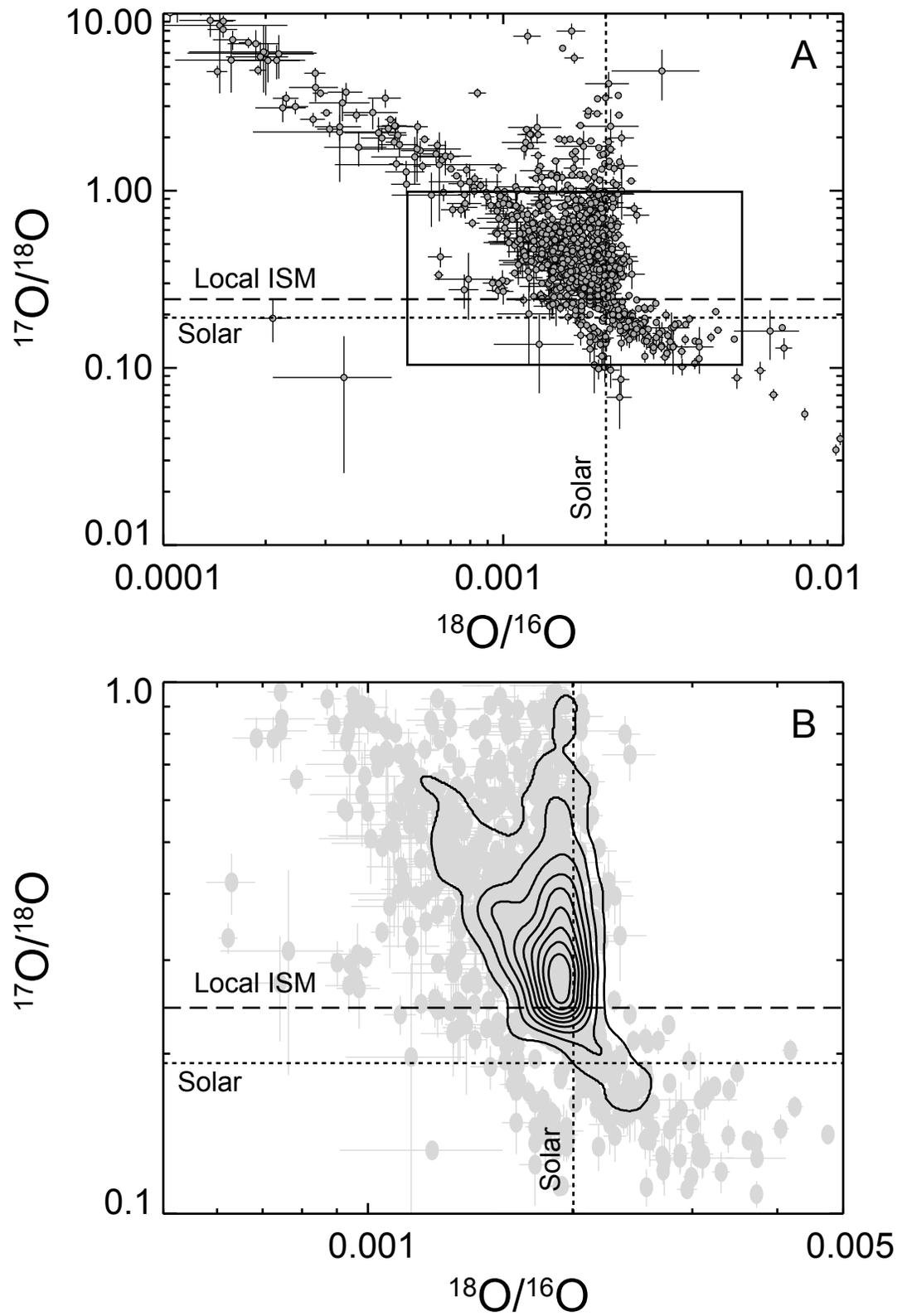

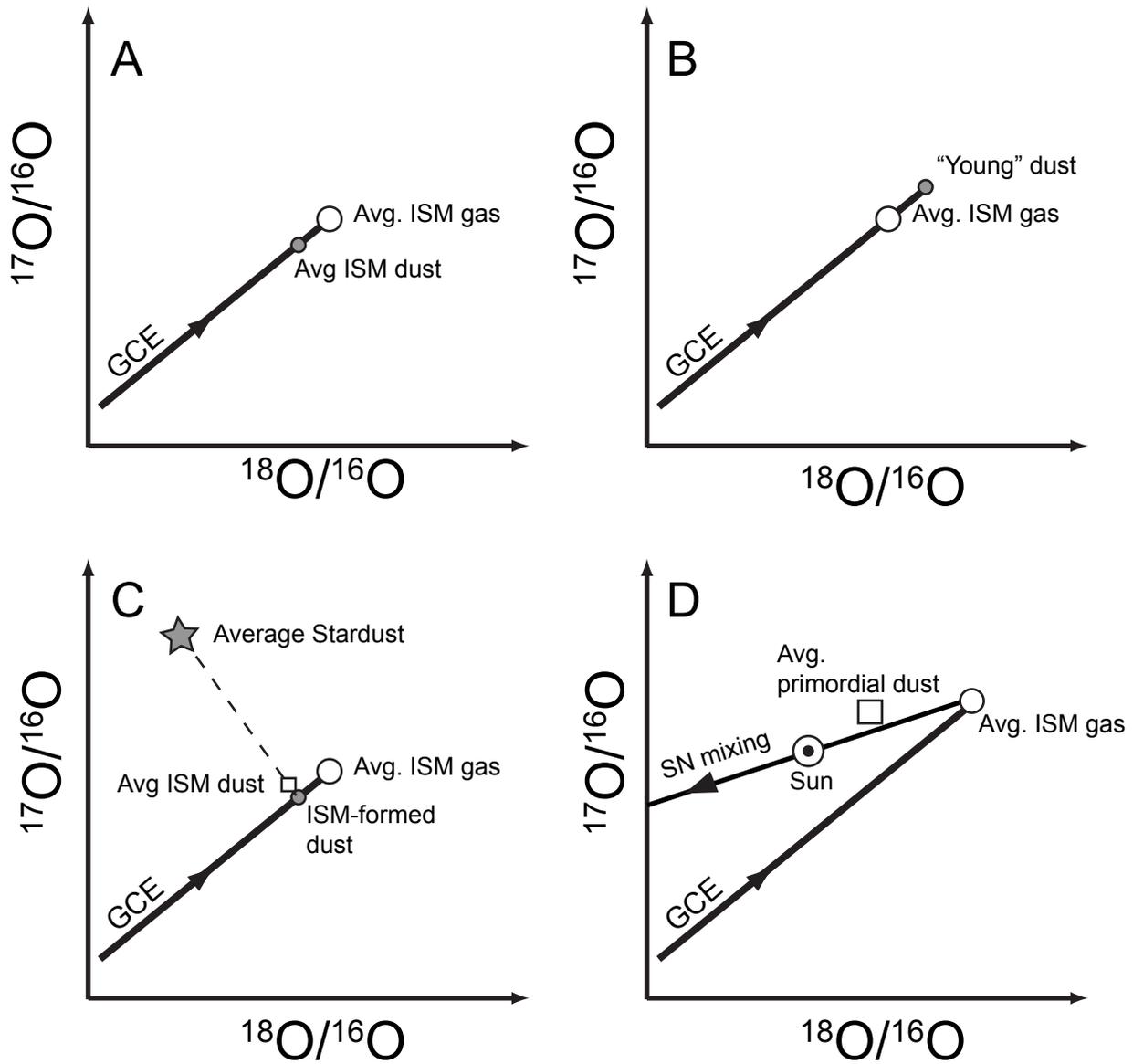

Figure 11

# Figure 12

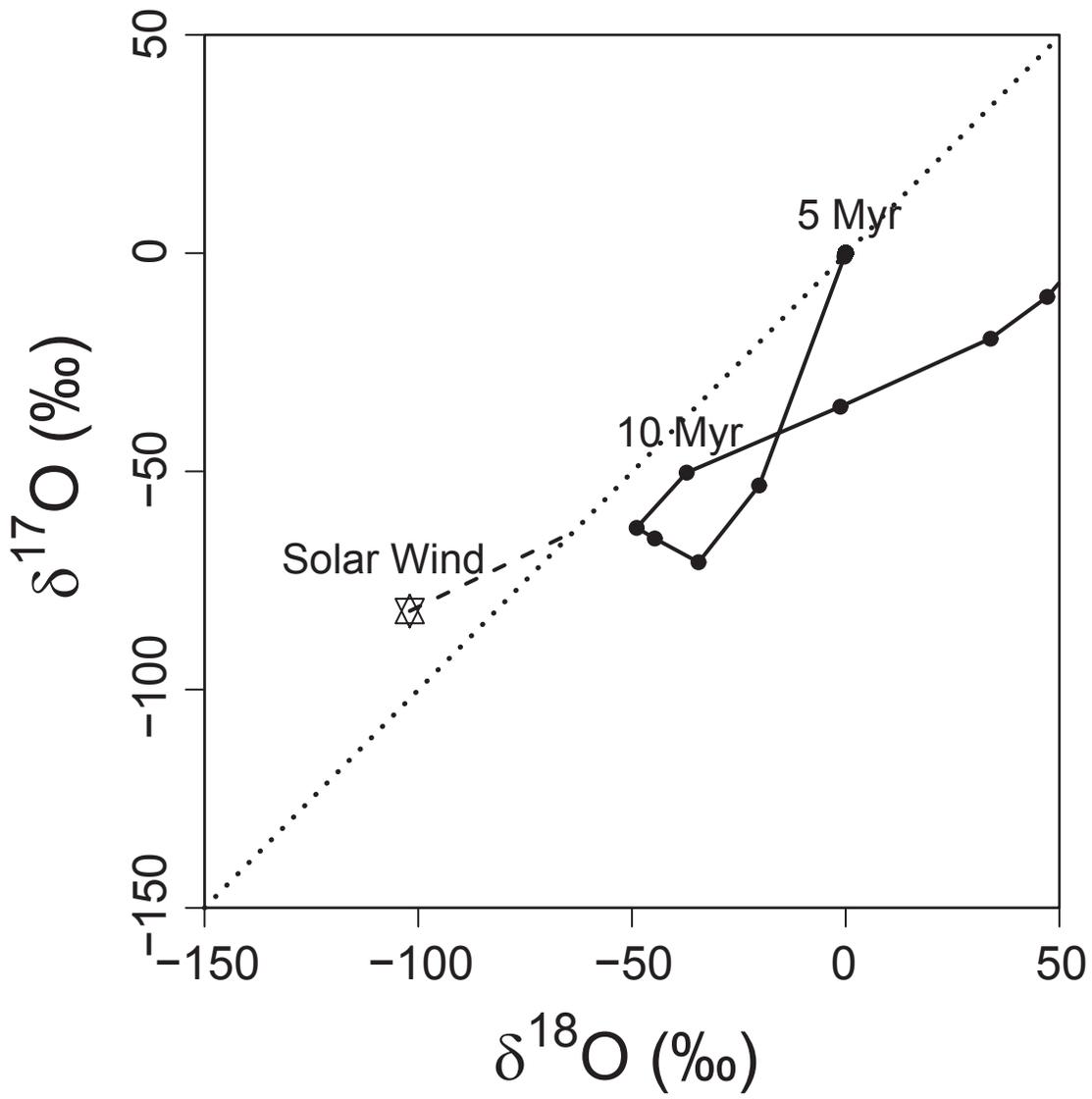